%% file: nnllfast-v2.tex
\documentclass[11pt,a4paper,numbers=noenddot]{scrartcl}

\usepackage[utf8]{inputenc}
\usepackage[T1]{fontenc}
\usepackage{lmodern}

\usepackage[english]{babel}

\usepackage{graphicx}
\usepackage{epsfig}
\usepackage{float}
\usepackage{cite}
\usepackage{amssymb}
\usepackage{amsmath}
\usepackage{url}
\usepackage[dvipsnames]{xcolor}
\usepackage{cancel}
\usepackage{nicefrac}
\usepackage[normalem]{ulem}
\usepackage[onehalfspacing]{setspace}
\usepackage{comment}
\usepackage{booktabs}
\usepackage{enumitem}
\usepackage{array}

\usepackage{listings}
\definecolor{codegreen}{rgb}{0,0.6,0}
\definecolor{codegray}{rgb}{0.5,0.5,0.5}
\definecolor{codepurple}{rgb}{0.58,0,0.82}
\definecolor{backcolour}{rgb}{0.95,0.95,0.92}

\lstdefinestyle{mystyle}{
    backgroundcolor=\color{backcolour},   
    commentstyle=\color{codegreen},
    keywordstyle=\color{magenta},
    numberstyle=\tiny\color{codegray},
    stringstyle=\color{codepurple},
    basicstyle=\ttfamily,
    columns=flexible,
    breakatwhitespace=false,
    breaklines=true,
    keepspaces=true,
    showspaces=false,
    showstringspaces=false,
    showtabs=false,
    tabsize=1,
    frame=single,
    belowskip=0pt
}
\usepackage[format=plain,labelfont={bf},font={small}]{caption}

\usepackage[colorlinks=true,citecolor=blue,linkcolor=blue]{hyperref}

\numberwithin{equation}{section}

\setkomafont{disposition}{\normalcolor\normalfont\bfseries}

\newcommand{\email}[1]{\href{mailto:#1}{\nolinkurl{#1}}}

\newcommand{\nnllfast}{\textsc{NNLL-fast}}
\newcommand{\prospino}{\textsc{Prospino}}
\newcommand{\nnlonnll}{NNLO$_\text{Approx}$+NNLL}

\newcommand{\alphas}{\alpha_{\mathrm{s}}}
\newcommand{\mav}{m_{\mathrm{av}}}
\newcommand{\msq}{m_{\tilde q}}
\newcommand{\mst}{m_{\tilde t_1}}
\newcommand{\mgl}{m_{\tilde g}}
\newcommand{\glu}{{\tilde g}}
\newcommand{\squ}{{\tilde q}}
\newcommand{\sto}{{\tilde t}}
\newcommand{\sbo}{{\tilde b}}

\begin{document}

\renewcommand*{\thefootnote}{\fnsymbol{footnote}}

\begin{flushright}
	\small KA-TP-07-2024\\
	\small MS-TP-24-08
\end{flushright}
\begin{center}
	{\LARGE \bfseries NNLL-fast 2.0: Coloured Sparticle Production at the LHC Run 3 with \boldmath$\sqrt{S} = 13.6$ TeV \par}

	\vspace{.7cm}
    Wim Beenakker\textsuperscript{a,}\footnote{\email{w.beenakker@science.ru.nl}},
	Christoph Borschensky\textsuperscript{b,}\footnote{\email{christoph.borschensky@kit.edu}},
    Michael Kr\"amer\textsuperscript{c,}\footnote{\email{mkraemer@physik.rwth-aachen.de}},
	Anna Kulesza\textsuperscript{d,}\footnote{\email{anna.kulesza@uni-muenster.de}},
    Eric Laenen\textsuperscript{e,f,g,}\footnote{\email{eric.laenen@nikhef.nl}},
    Judita Mamužić\textsuperscript{h,}\footnote{\email{judita.mamuzic@cern.ch}},
	Laura Moreno Valero\textsuperscript{d,}\footnote{\email{l_more02@uni-muenster.de}}

	\vspace{.3cm}
	\textit{
        \textsuperscript{a }Institute for Mathematics, Astrophysics and Particle Physics, Radboud University Nijmegen, Heyendaalseweg 135, Nijmegen, The Netherlands\\[.2em]
		\textsuperscript{b }Institute for Theoretical Physics, Karlsruhe Institute of Technology, Wolfgang-Gaede-Str.\ 1, 76131 Karlsruhe, Germany\\[.2em]
        \textsuperscript{c }Institute for Theoretical Particle Physics and Cosmology, RWTH Aachen University, Sommerfeldstr.\ 16, 52074 Aachen, Germany\\[.2em]
		\textsuperscript{d }Institut f\"ur Theoretische Physik, University of M\"unster, Wilhelm-Klemm-Str.\ 9, 48149 M\"unster, Germany\\[.2em]
        \textsuperscript{e }Institute of Physics/Institute for Theoretical Physics Amsterdam, University of Amsterdam, Science Park 904, 1098 XH Amsterdam, The Netherlands\\[.2em]
        \textsuperscript{f }Nikhef, Theory Group, Science Park 105, 1098 XG, Amsterdam, The Netherlands\\[.2em]
        \textsuperscript{g }Institute for Theoretical Physics, Utrecht University, Leuvenlaan 4, 3584 CE Utrecht, The Netherlands\\[.2em]
        \textsuperscript{h }Institut de Física d'Altes Energies (IFAE), Edifici Cn, Campus UAB, 08193 Bellaterra (Barcelona), Spain
	}
\end{center}

\thispagestyle{empty}
\vfill
\pagebreak

\renewcommand*{\thefootnote}{\arabic{footnote}}
\setcounter{footnote}{0}

\vspace*{0.1cm}
\centerline{\bfseries Abstract}
\begin{abstract}
	We report on updated precision predictions for total cross sections of coloured supersymmetric particle production at the LHC with a centre-of-mass energy of $\sqrt S = 13.6$~TeV, computed with the modern PDF4LHC21 set. The cross sections are calculated at an approximated NNLO accuracy in QCD and contain corrections from the threshold resummation of soft-gluon emission up to NNLL accuracy as well as Coulomb-gluon contributions including bound-state terms. The corrections are found to increase the cross sections and reduce the theoretical uncertainty as compared to the best available fixed-order calculations. These predictions constitute the state-of-the-art calculations and update the existing results for $\sqrt S = 13$~TeV. We make our new results publicly available in the version 2.0 update to the code package \nnllfast{}.
\end{abstract}

\tableofcontents

\input{1-introduction}
\input{2-squgluproduction}
\input{3-hocalculations}
\input{4-nnllfast}
\input{5-numresults}
\input{6-conclusions}

%%%%%%%%%%%%%%%%%%%%%%%%%%%%%%%%%%%%%%%%%%%
\section*{Acknowledgements}
%%%%%%%%%%%%%%%%%%%%%%%%%%%%%%%%%%%%%%%%%%%
We are very grateful to our collaborators who have contributed to the \textsc{NLL-fast} and \nnllfast{} projects over the years: S.\ Brensing-Thewes, R.\ Heger, M.\ Mangano, S.\ Marzani, L.\ Motyka, I.\ Niessen, J.\ Rojo, S.\ Padhi, T.\ Plehn, X.\ Portell, D.\ Schwartländer, and V.\ Theeuwes. MK acknowledges support from the DFG under grant 396021762 - TRR 257: Particle Physics Phenomenology after the Higgs Discovery. The work of LMV  was supported by the DFG Research Training Group “GRK 2149: Strong and Weak Interactions - from Hadrons to Dark Matter”.

\bibliographystyle{spphys_custom}
\bibliography{bibliography}

\end{document}

%% file: 1-introduction.tex
%%%%%%%%%%%%%%%%%%%%%%%%%%%%%%%%%%%%%%%%%%%
\section{Introduction}\label{s:intro}
%%%%%%%%%%%%%%%%%%%%%%%%%%%%%%%%%%%%%%%%%%%

The search for supersymmetry (SUSY) \cite{Wess:1973kz,Wess:1974tw,Fayet:1976et,Farrar:1978xj,Sohnius:1985qm,Martin:1997ns} is one of the most important objectives of the physics programme of the Large Hadron Collider (LHC).  SUSY addresses the shortcomings of the Standard Model (SM) of particle physics in an elegant and compelling way. Consequently,  over many decades, and in particular since the beginning of the LHC operations, there has been an immense interest of the community in the results of  the searches. SUSY posits that each elementary particle in the SM is paired to a supersymmetric partner or sparticle $-$ with squarks ($\tilde{q}$) and gluinos ($\tilde{g}$) being the superpartners of quarks and gluons, respectively. In the context of the Minimal Supersymmetric Standard Model (MSSM) with $R$-parity conservation \cite{Nilles:1983ge,Haber:1984rc}, one of the most studied SUSY models, sparticles are always produced in pairs. Exactly such production processes have been and are currently being searched for at the LHC. On the basis of Run 1 and 2 data analyses at the ATLAS \cite{ATLAS:1999uwa} and CMS \cite{CMS:2006myw} experiments the bounds on coloured sparticles masses of up to 1--1.9~TeV for squarks and 1.2--2.5~TeV for gluinos have been determined, with exact values depending on additional mass parameters of the electroweak SUSY sector and the examined search channel \cite{ATLAS:2017tmw,ATLAS:2018nud,ATLAS:2020syg,ATLAS:2021kxv,ATLAS:2022ckd,ATLAS:2022ihe,ATLAS:2023afl,CMS:2019zmd,CMS:2019ybf,CMS:2020cur,CMS:2020bfa,CMS:2021beq,CMS:2023xlp,CMS:2023zuu}. For the third generation squarks, i.e.\ the stops and sbottoms as superpartners of the top and bottom quarks, the experimental limits are a bit more relaxed, excluding stops and sbottoms up to masses of around 0.5--1.6~TeV, depending on the search channel \cite{ATLAS:2019gdh,ATLAS:2020dsf,ATLAS:2020xzu,ATLAS:2021yij,ATLAS:2021kxv,ATLAS:2021pzz,ATLAS:2021jyv,ATLAS:2023dbq,CMS:2019ybf,CMS:2021eha,CMS:2023ktc,CMS:2023yzg}.

One of the very important ingredients entering the experimental analysis and enabling an accurate derivation of the mass exclusion limits are precise theoretical predictions for the total cross sections for the processes of interest. The next-to-leading order (NLO) SUSY-QCD corrections to squark and gluino production, both for total production rates, decays, as well as differential distributions, have been calculated some time ago \cite{Beenakker:1994an,Beenakker:1995fp,Beenakker:1996ch,Beenakker:1997ut,Hollik:2012rc,Goncalves-Netto:2012nvl,Hollik:2013xwa,Gavin:2013kga,Gavin:2014yga,Degrande:2015vaa,Frixione:2019fxg}. The electroweak NLO corrections are also known \cite{Hollik:2007wf,Beccaria:2008mi,Hollik:2008yi,Hollik:2008vm,Mirabella:2009ap,Germer:2010vn,Germer:2014jpa,Hollik:2015lha}. Due to the high mass limits, the kinematical region where squarks and gluinos are produced close to their production threshold is of increased importance, and a significant contribution to the total cross section comes from this region. Near threshold, additional hard gluon radiation is strongly suppressed, forcing the radiation to be soft. Soft radiation, in turn, brings about large logarithmic contributions to the cross sections, which need to be systematically taken into account. The summation of the soft-gluon contributions to all orders in the strong coupling constant $\alphas$ can be performed by means of threshold resummation techniques in Mellin-moment space \cite{Sterman:1986aj,Catani:1989ne,Bonciani:1998vc,Contopanagos:1996nh,Kidonakis:1998bk,Kidonakis:1998nf}. Resummed results for squark and gluino production, including stops, were first obtained at the next-to-leading logarithmic (NLL) accuracy, both in the Mellin-space approach \cite{Kulesza:2008jb,Kulesza:2009kq,Beenakker:2009ha,Beenakker:2010nq,Beenakker:2011fu,Beenakker:2011dk,Borschensky:2014cia,Beenakker:2015rna} and in the framework of soft collinear effective theory (SCET) \cite{Beneke:2009rj,Beneke:2010da,Falgari:2012hx}. The accuracy of resummation was later increased to the next-to-next-to-leading logarithmic (NNLL) level, again in both the Mellin-space approach \cite{Beenakker:2011sf,Langenfeld:2012ti,Pfoh:2013iia,Beenakker:2013mva,Beenakker:2014sma,Beenakker:2016gmf,Beenakker:2016lwe} and in SCET \cite{Beneke:2013opa,Broggio:2013cia,Beneke:2014wda,Beneke:2016kvz}. Recently, in \cite{Borschensky:2024zdg}, soft-gluon corrections in the Mellin-space resummation formalism have also been calculated for squark production in a non-minimal SUSY model, the Minimal R-symmetric Supersymmetric Standard Model \cite{Kribs:2007ac}, and matched to the existing NLO-QCD corrections \cite{Diessner:2017ske}.

In this work, we report on updated predictions for the cross sections for squark and gluino production processes in the MSSM at the approximated next-to-next-to-leading-order (NNLO) matched to NNLL accuracy for LHC Run 3 with a collision energy of $\sqrt{S} = 13.6$~TeV. The \nnlonnll{} results are the most precise theoretical predictions currently available, including also resummation of Coulomb contributions as well as corrections from bound-state formation in the final state. The results have been consistently used by both the ATLAS and CMS collaborations in the analyses of SUSY searches in Run~2. The predictions for Run~3, presented here, can be obtained with the version 2.0 of the publicly available package \nnllfast{}. They correspond to an update of the Run~2 predictions in \cite{Beenakker:2016lwe}, provided by earlier versions of the package, in line with the upgrade at the LHC Run~3. The two sets of predictions  differ not only by the value of the centre-of-mass energy but also by the sets of parton distribution functions (PDFs) with which they are obtained. The aim of this paper, similarly to~\cite{Borschensky:2014cia} for NLO+NLL calculations, is to provide in one document a brief overview of the results that can be obtained with \nnllfast{}~2.0 (central values of the cross sections, error estimates and the $K$-factors), together with the calculations that led to them, as well as to discuss the impact of the differences in the \nnllfast{} set-up on the predictions.

The paper is structured as follows. In Sec.~\ref{s:squgluprod} we introduce the production processes of interest for this work. In Sec.~\ref{s:hocalc}, we review the higher-order calculations and in particular the threshold resummation formalism at NNLL accuracy, and briefly discuss the treatment of the various uncertainties. The implementation and parameters used in the code package \nnllfast{} as well as the version 2.0 update is detailed in Sec.~\ref{s:nnllfast}. Numerical results are presented in Sec.~\ref{s:numerics}, where we also provide comparisons with results obtained using an earlier version of the \nnllfast{} code, and we conclude in Sec.~\ref{s:conclusion}.

%% file: 2-squgluproduction.tex
%%%%%%%%%%%%%%%%%%%%%%%%%%%%%%%%%%%%%%%%%%%
\section{Squark and gluino production at the LHC}\label{s:squgluprod}
%%%%%%%%%%%%%%%%%%%%%%%%%%%%%%%%%%%%%%%%%%%
The coloured sector of the MSSM consists of the superpartners of the quarks and gluons, the scalar squarks $\squ$ and the fermionic gluinos $\glu$, the latter of which are of Majorana fermionic nature. Due to the colour charge of squark and gluinos, their production cross sections at the LHC are predicted to be large and dominate over cross sections for other supersymmetric particles. This results in the already mentioned relatively high exclusion limits on the squark and gluino masses (in comparison with exclusion limits on masses of other particles), established from Run 1 and 2 data. The corresponding experimental analysis relies often on certain simplified scenarios such as decoupling limits where all supersymmetric particles other than the ones that are searched for are assumed to be very heavy and therefore decoupled from the production process and out of reach for direct searches with current experiments.

\begin{figure}[tp]
	\centering
	\begin{tabular}{m{3mm} m{11cm}}
		\textbf{(a)} & \includegraphics[width=0.25\textwidth]{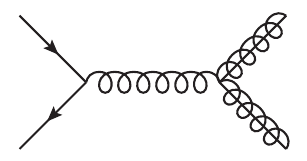}\includegraphics[width=0.25\textwidth]{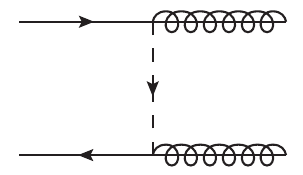}\includegraphics[width=0.25\textwidth]{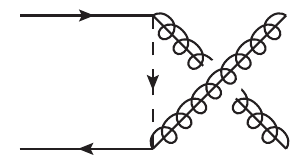}\newline
		\includegraphics[width=0.25\textwidth]{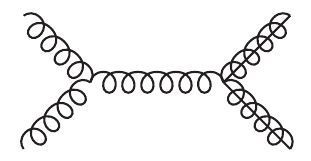}\includegraphics[width=0.25\textwidth]{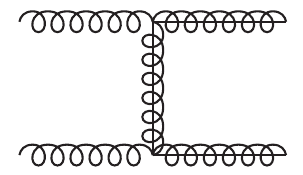}\includegraphics[width=0.25\textwidth]{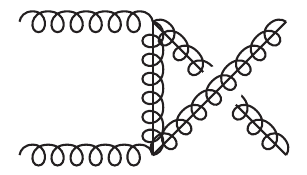} \\\midrule
		\textbf{(b)} & \includegraphics[width=0.25\textwidth]{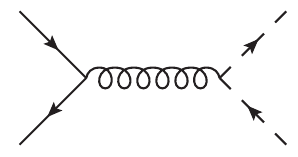}\includegraphics[width=0.25\textwidth]{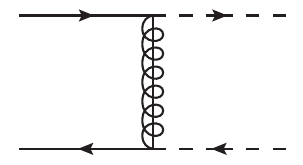}\includegraphics[width=0.25\textwidth]{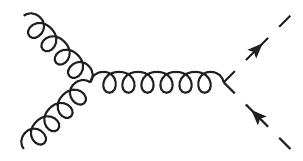}\newline
		\includegraphics[width=0.25\textwidth]{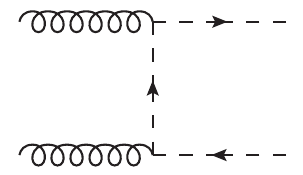}\includegraphics[width=0.25\textwidth]{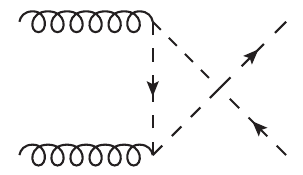}\includegraphics[width=0.25\textwidth]{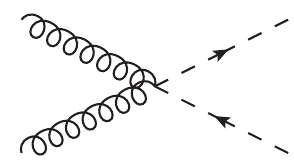} \\\midrule
		\textbf{(c)} & \includegraphics[width=0.25\textwidth]{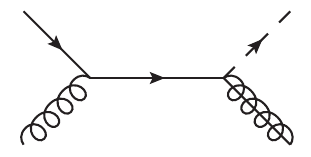}\includegraphics[width=0.25\textwidth]{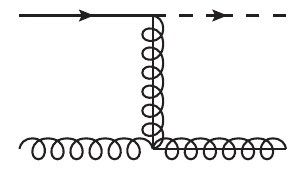}\includegraphics[width=0.25\textwidth]{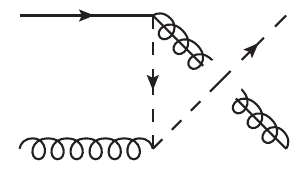} \\\midrule
		\textbf{(d)} & \includegraphics[width=0.25\textwidth]{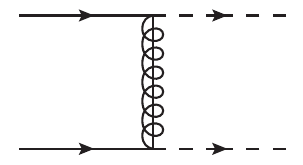}\includegraphics[width=0.25\textwidth]{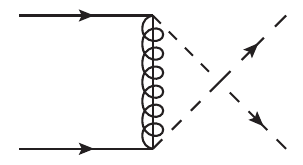}
	\end{tabular}
	\caption{Tree-level Feynman diagrams for the partonic squark and gluino production processes of Eq.~\eqref{eq:pprodchannels}: (a) gluino-pair production $\glu\glu$, (b) squark-antisquark production $\squ\squ^*$, (c) squark-gluino production $\squ\glu$, (d) squark-pair production $\squ\squ$. Solid lines with an arrow correspond to quarks, curly lines to gluons, dashed lines with an arrow to squarks, and curly lines with a solid line in the middle to gluinos.}
	\label{fig:losqglproduction}
\end{figure}

Assuming $R$-parity conservation, supersymmetric particles can only be produced in pairs. In the following, we will discuss only the dominant SUSY-QCD production channels. For squarks and gluinos, the following inclusive pair production processes can take place at a hadron collider with two colliding hadrons $h_1$ and $h_2$ (where in the case of the LHC, $h_1$ and $h_2$ are both protons):
\begin{equation}\label{eq:hprodchannels}
	h_1 h_2 \to \glu\glu,\, \squ\squ^*,\, \squ\glu,\, \squ\squ\, + X\,,
\end{equation}
where $X$ stands for any additional radiation. We label the four types of processes in the following as:
\begin{itemize}
	\item $\glu\glu$: \emph{gluino-pair} production,
	\item $\squ\squ^*$: \emph{squark-antisquark} production,
	\item $\squ\glu$: \emph{squark-gluino} production,
	\item $\squ\squ$: \emph{squark-squark} or \emph{squark-pair} production.
\end{itemize}
For the latter two processes, here and in the following, we always assume the charge-conjugated processes $h_1 h_2 \to \squ^*\glu,\, \squ^*\squ^*\, + X$ to be implied\footnote{Note that in the MSSM, as gluinos are Majorana fermions, they are their own antiparticles, such that we do not distinguish between $\glu$ and $\bar\glu$.}, i.e.\ when e.g.\ referring to $\squ\squ$ production, we mean the sum of $\squ\squ$ and $\squ^*\squ^*$. At the partonic level, the following initial-state channels contribute to the production processes at leading order (LO):
\begin{equation}\label{eq:pprodchannels}
	\begin{split}
		q_i\bar q_i,\, gg \to \glu\glu\,,\qquad q_{i'}\bar q_{j'},\, gg \to \squ_i\squ_j^*\,,\qquad q_i g \to \squ_i \glu\,,\qquad q_i q_j \to \squ_i\squ_j\,,
	\end{split}
\end{equation}
as well as the charge-conjugated processes, whenever appropriate. Here, the (s)quark indices $i^{(\prime)}$ and $j^{(\prime)}$ denote the (s)quark flavour. While squark-antisquark production through the $gg$ initial-state channel is always flavour diagonal, squark-antisquark and squark-pair final states include also squarks of different flavours \mbox{$i\ne j$}, produced through the $q\bar q$ and $qq$ initial-state channels, respectively. The tree-level Feynman diagrams for the partonic production channels in Eq.~\eqref{eq:pprodchannels} are shown in Figure~\ref{fig:losqglproduction}.

In Eq.~\eqref{eq:hprodchannels}, we sum over the two chirality states $\squ_L$, $\squ_R$ of the squarks\footnote{As squarks are scalar particles, they cannot carry chirality. The labels $L$ and $R$ are only used to distinguish the superpartners of left- and right-handed quarks. We furthermore neglect the masses of the five light quarks other than the top quark, so that the $L$ and $R$ squark states correspond to their mass eigenstates.}. We assume the superpartners of the light quarks $(u, d, c, s, b)$ to be mass degenerate, leading to a 10-fold squark degeneracy. Due to the absence of top quark densities in parton distribution functions (PDF), there are fewer diagrams  for the production of superpartners of the top quark than for the production of supersymmetric partners of light quarks (which e.g.\ includes the second diagram with $t$-channel gluino exchange in  Figure~\ref{fig:losqglproduction}~(b)). Moreover, also in contrast to the light quark case, mixing effects of the left- and right-handed superpartners in the stop mass matrix cannot be neglected due to the large top quark masses. We thus consider stop-antistop production separately:
\begin{equation}
	h_1 h_2 \to \sto_a\sto_a^*\, + X\qquad a = 1, 2\,,
\end{equation}
where $a = 1, 2$ conventionally denote the light and heavy states of the stop, respectively\footnote{We do not consider mixed $\sto_1\sto_2^*$ or same-charge $\sto_{1,2}\sto_{1,2}$ production, as these processes are strongly suppressed at tree-level by vanishing top quark PDFs and thus would have to be considered as loop-induced processes, which receive the usual suppression from small coupling constants and loop factors with respect to the above-mentioned tree-level processes.}. When appropriate, the same treatment can be applied also to the case of the superpartners of the bottom quark, the sbottoms, so that we consider their production separately, and thus only assume an 8-fold degeneracy of the superpartners $\squ$ of $(u, d, c, s)$.

The hadronic production cross section for squark and gluino production can be written as a convolution of PDFs and the partonic cross section:
\begin{equation}\label{eq:hadr-cross}
	\begin{split}
		\sigma_{h_1 h_2 \to kl}\bigl(\rho, \{m^2\}\bigr) &= \sum_{i,j} \int dx_1\, dx_2\,d\hat{\rho}\,\delta\!\left(\hat{\rho} - \frac{\rho}{x_1 x_2}\right)\\
		&\qquad\qquad\times f_{i/h_1}(x_1, \mu^2)\, f_{j/h_2}(x_2, \mu^2)\, \sigma_{ij \to kl}\bigl(\hat{\rho},\{ m^2\},\mu^2\bigr)\,,
	\end{split}
\end{equation}
with $k, l = \squ^{(*)}, \glu, \sto^{(*)}$ and $i, j = q, \bar q, g$. Here, the variable $\rho := (m_k + m_l)^2/S$ is given by the ratio of the sum of the final-state masses $m_k$ and $m_l$ squared with respect to the hadronic squared centre-of-mass energy $S$. $\{m^2\}$ stands for all the masses (such as squark and gluino masses) entering the calculation. Furthermore, $f_{i/h_1}(x_1, \mu^2)$ and $f_{j/h_2}(x_2, \mu^2)$ denote the PDFs which can, at LO, be interpreted as probabilities for partons with flavours $i$ and $j$ to be present inside the hadrons $h_1$ and $h_2$ and carrying momentum fractions $x_1$ and $x_2$, respectively, of the full hadronic momenta. $\sigma_{ij \to kl}$ stands for the partonic cross section. The scale $\mu$ corresponds to the factorisation scale $\mu_F$, separating long- and short-distance physics, which we set equal to the renormalisation scale in all of our calculations, $\mu := \mu_F = \mu_R$.

%% file: 3-hocalculations.tex
%%%%%%%%%%%%%%%%%%%%%%%%%%%%%%%%%%%%%%%%%%%
\section{Higher-order calculations}\label{s:hocalc}
%%%%%%%%%%%%%%%%%%%%%%%%%%%%%%%%%%%%%%%%%%%
Since the considered production processes are dominated by QCD and SUSY-QCD interactions, higher-order corrections can be sizeable. Therefore, they need to be taken into account in order to obtain reliable theoretical predictions and to reduce theoretical uncertainties. Below we briefly review the calculations for current state-of-the-art \nnlonnll{} predictions for inclusive cross sections for squark and gluino production at the LHC.

%%%%%%%%%%%%%%%%%%%%%%%%%%%%%%%%%%%%%%%%%%%
\subsection{Fixed-order contribution}
%%%%%%%%%%%%%%%%%%%%%%%%%%%%%%%%%%%%%%%%%%%
The fixed-order contribution NNLO$_\text{Approx}$ is an approximation of the NNLO SUSY-QCD result, consisting of  $\sigma_{h_1 h_2 \to kl}^{\mathrm{NLO}}$, the full NLO SUSY-QCD cross section at $\mathcal{O}(\alphas^3)$, and $\Delta\sigma_{h_1 h_2 \to kl}^{\mathrm{NNLO_{Approx}}}$  which is an approximation of the $\mathcal{O}(\alphas^4)$ corrections:
\begin{equation}\label{eq:nlo-nnloapprox}
	\sigma_{h_1 h_2 \to kl}^{\mathrm{NNLO_{Approx}}} = \sigma_{h_1 h_2 \to kl}^{\mathrm{NLO}} + \Delta\sigma_{h_1 h_2 \to kl}^{\mathrm{NNLO_{Approx}}}\,.
\end{equation}
The inclusive NLO SUSY-QCD production cross sections for squarks and gluinos at hadron colliders have been calculated over 25 years ago \cite{Beenakker:1996ch,Beenakker:1997ut}, and are implemented in the \prospino{} code \cite{Beenakker:1996ed}, where in the more recent \prospino{}~2 version additional SUSY processes have been included. More recently, about 10 years ago, the calculation of squark and gluino production at NLO-QCD has been automated based on the \textsc{MadGolem} tool \cite{Goncalves-Netto:2012nvl}, and the squark-antisquark and squark-pair production processes have been recalculated at NLO-QCD including also decays and matching to parton showers, keeping all squark masses separate, i.e.\ without assuming a 10-fold squark degeneracy \cite{Gavin:2013kga,Gavin:2014yga}. Squark and gluino production can now be calculated, both for differential and total rates, in a fully automatised manner up to NLO-QCD using the \textsc{MadGraph5\_aMC\@NLO} tool \cite{Frixione:2019fxg}. We refer the reader interested in the details of the NLO calculations to the original literature.

The $\Delta\sigma_{h_1 h_2 \to kl}^{\mathrm{NNLO_{Approx}}}$ correction collects the $\mathcal{O}(\alphas^4)$ contributions which are enhanced in the limit of sparticle pair-production taking place close to the threshold, originating from soft-gluon radiation, Coulomb-like emissions as well as two-loop non-Coulomb potential and kinetic-energy corrections, see \cite{Beneke:2009ye}. Correspondingly, special care needs to be taken when matching $\Delta\sigma_{h_1 h_2 \to kl}^{\mathrm{NNLO_{Approx}}}$ and the full NLO correction to the resummed results, as discussed in the next section.

%%%%%%%%%%%%%%%%%%%%%%%%%%%%%%%%%%%%%%%%%%%
\subsection{Threshold resummation}\label{s:hocalc:threshold}
%%%%%%%%%%%%%%%%%%%%%%%%%%%%%%%%%%%%%%%%%%%
Due to the high exclusion limits, squarks and gluinos -- should they exist -- must be heavy, and the dominant contribution to their production cross sections stems from the threshold region where the sum of the final-state masses is close to the hadronic centre-of-mass energy, $S \to (m_k + m_l)^2$. In this limit, all additional radiation at higher orders is constrained to be soft, and the corrections due to soft-gluon emission have the general form
\begin{equation}\label{eq:beta}
	\alphas^n \ln^m\!\beta^2\,, \ m\leq 2n \qquad \text{ with } \qquad \beta^2 \equiv 1 - \hat{\rho} = 1 - \frac{4\mav^2}{s}\,,
\end{equation}
where $\mav := (m_k + m_l)/2$ is the average mass of the final-state particles $k$ and $l$, $s=x_1x_2S$ is the partonic centre-of-mass energy squared, and $\alphas$ denotes the strong coupling. In the threshold limit, $\beta \to 0$, the logarithms of Eq.~\eqref{eq:beta} become large and thus have to be taken into account at all orders not to spoil the perturbative expansion in $\alphas$.

We carry out the all-order resummation of the soft-gluon emission after taking a Mellin transform of the hadronic cross section,
\begin{equation}\label{eq:Mellin-transf}
	\begin{split}
		\tilde\sigma_{h_1 h_2 \to kl}\bigl(N, \{m^2\}\bigr) &\equiv \int_0^1 d\rho\;\rho^{N-1}\;\sigma_{h_1 h_2\to kl}\bigl(\rho,\{ m^2\}\bigr)\\
		&= \;\sum_{i,j} \,\tilde f_{i/{h_1}} (N+1,\mu^2)\,\tilde f_{j/{h_2}} (N+1, \mu^2) \,\tilde{\sigma}_{ij \to kl}\bigl(N,\{m^2\},\mu^2\bigr)\,,
	\end{split}
\end{equation}
where the threshold-enhanced terms are now of the form $\alphas^n \log^m\!N$, $m\leq 2n$, depending on the Mellin moments $N$, and the threshold limit is given by $N\to \infty$. The \mbox{$\tilde f_{i/{h_1}} (N+1, \mu^2)$} denote the PDFs in Mellin space. The Mellin-transformed partonic cross section $\tilde{\sigma}_{ij \to kl}\bigl(N,\{m^2\},\mu^2\bigr)$ then factorises into a product of terms separating hard and soft as well as soft-collinear contributions, allowing for a systematic reorganisation of the enhanced logarithms in terms of exponential functions \cite{Sterman:1986aj,Catani:1989ne,Bonciani:1998vc,Contopanagos:1996nh,Kidonakis:1998bk,Kidonakis:1998nf}. The fully factorised result in terms of the resummed functions is then given as:
\begin{equation}\label{eq:resummed-cross}
	\begin{split}
		\tilde{\sigma}^{\mathrm{(res)}}_{ij\to kl}\bigl(N,\{m^2\},\mu^2\bigr) &= \sum_{I}\,\tilde\sigma^{(0)}_{ij\to kl,I}\bigl(N,\{m^2\},\mu^2\bigr)\, C_{ij\to kl,I}(N,\{m^2\},\mu^2)\\
		&\qquad \times\,\Delta_i\Delta_j\Delta^{\mathrm{(s)}}_{ij\to kl,I} (N+1,Q^2,\mu^2)\,,
	\end{split}
\end{equation}
where we introduced the hard scale $Q^2 = 4\mav^2$, and where the cross section is split up into colour channels $I$ in an $s$-channel colour basis, in which the factorisation of soft and hard parts becomes diagonal \cite{Kulesza:2008jb,Kulesza:2009kq,Beneke:2009rj}, and $\Delta_i\Delta_j\Delta^{\mathrm{(s)}}_{ij\to kl,I}$ are the functions containing the resummed threshold logarithms:
\begin{equation}\label{eq:NNLL-expa}
	\Delta_i\Delta_j\Delta^{\mathrm{(s)}}_{ij\to kl,I} \;=\; \exp\Big[L g_1(\alphas L) + g_2(\alphas L) + \alphas g_3(\alphas L) + \ldots \Big]
\end{equation}
with $L := \ln N$. In Eq.~\eqref{eq:NNLL-expa}, the perturbative series is now organised differently: while the exponential function takes into account terms up to all orders in $\alphas$, the functions $g_1$, $g_2$, $g_3$ etc.\ now define different  logarithmic orders of the approximation, with the first summand in the exponent $Lg_1(\alphas L)$ resumming terms up to leading-logarithmic (LL) accuracy, including additionally the second term $g_2(\alphas L)$ denotes next-to-leading logarithmic (NLL) and including also the third term $\alphas g_3(\alphas L)$ denotes next-to-next-to-leading logarithmic (NNLL) accuracy of the threshold resummation. Expressions for the $g_1$, $g_2$, and $g_3$ functions can be found in e.g.\ \cite{Kulesza:2009kq,Beenakker:2011sf}.

In Eq.~\eqref{eq:resummed-cross}, the matching coefficient $C_{ij\to kl, I}(N, \{m^2\}, \mu^2)$ is given by:
\begin{equation}\label{eq:factCcoeff}
	C_{ij\to kl, I} = \mathcal{C}_{ij\to kl, I}^{\text{Coulomb}} \times (1 + \frac{\alphas}{\pi}\,\mathcal{C}_{ij\to kl, I}^{(1)} + \dots)\,.
\end{equation}
The factor $\mathcal{C}_{ij\to kl, I}^{\text{Coulomb}}$ in Eq.~\eqref{eq:factCcoeff} resums threshold-enhanced terms due to Coulomb-gluon exchange between slowly-moving final-state particles by employing the Coulomb Green’s function of non-relativistic QCD with a NLO Coulomb potential, see \cite{Beenakker:2016lwe} for more details. The terms that are non-logarithmic in $N$ from the NLO corrections, including one-loop virtual contributions, but excluding the Coulomb-gluon exchange between final states to avoid the double counting of these contributions, are collected by $\mathcal{C}_{ij\to kl, I}^{(1)}$.

In order to obtain physical results, the hadronic cross section in Mellin space must be transformed back to physical space. This is done by performing an inverse Mellin transform according to the minimal prescription \cite{Catani:1996yz}. In addition to the inverse Mellin transformation, we match the resummed cross section to the best available fixed-order calculation. To avoid the double counting of terms that occur in both the resummed as well as the fixed-order calculations, we expand the resummed cross section up to the available fixed order, which is of  $\mathcal{O}(\alphas^4)$ in our case, and subtract the expanded from the resummed part. Then, we add the fixed-order cross section $\sigma^{\mathrm{NNLO_{Approx}}}_{h_1 h_2 \to kl}$ of Eq.~\eqref{eq:nlo-nnloapprox}. We note that the threshold-enhanced two-loop contribution $\Delta\sigma_{h_1 h_2 \to kl}^{\mathrm{NNLO_{Approx}}}$ as included in $\sigma^{\mathrm{NNLO_{Approx}}}_{h_1 h_2 \to kl}$ differs from the corresponding term of the same order in the expansion of the resummed cross section by subleading contributions that are suppressed in Mellin space as $\mathcal{O}(1/N)$. The methodology of this matching procedure ensures that we combine the best known fixed-order result, covering also the kinematical region away from threshold, with the dominant threshold-enhanced corrections beyond the fixed order, yielding:
\begin{equation}\label{eq:matching}
	\begin{split}
		&\sigma^{\nnllfast}_{h_1 h_2 \to kl}\bigl(\rho, \{m^2\},\mu^2\bigr) = \sigma^{\mathrm{BS}}_{h_1 h_2\to kl}(\rho) + \sigma^{\mathrm{NNLO_{Approx}}}_{h_1 h_2 \to kl}\bigl(\rho,\{m^2\},\mu^2\bigr)\\
		&\qquad + \sum_{i,j}\int_\mathrm{CT}\frac{dN}{2\pi i}\;\rho^{-N}\,\tilde f_{i/h_1}(N+1,\mu^2)\,\tilde f_{j/h_{2}}(N+1,\mu^2)\\
		&\qquad\quad\times \left[\tilde\sigma^{\mathrm{(res,\,NNLL,\,Coulomb)}}_{ij\to kl}\bigl(N,\{m^2\},\mu^2\bigr) - \tilde\sigma^{\mathrm{(res,\,NNLL,\,Coulomb)}}_{ij\to kl}\bigl(N,\{m^2\},\mu^2\bigr){\left.\right|}_{\scriptscriptstyle{\mathrm{NNLO}}}\right]\,.
	\end{split}
\end{equation}
As explained above, we perform simultaneous resummation of soft-gluon emission up to NNLL and of threshold-enhanced Coulomb contribution, denoted by the superscripts `res,\,NNLL,\,Coulomb' in Eq.~\eqref{eq:matching}. In addition, corrections due to the formation of bound states between final-state particles are included in the $\sigma^{\mathrm{BS}}_{h_1 h_2\to kl}$ term. Again, we refer to the previous publication of \cite{Beenakker:2016lwe} for more details on the calculation of this term%
\footnote{The boundstate contributions are generally positive and have a moderate effect on the total cross sections, leading to an increase with respect to NLO of a few per mille to the per cent range, as shown in \cite{Beenakker:2016lwe}. The effects are the largest for processes with large colour factors such as $\glu\glu$, and they become more relevant close to threshold, i.e.\ for smaller centre-of-mass energies or larger final-state masses, see also \cite{Falgari:2012hx}.}%
. Our final result at \nnlonnll{} accuracy, denoting the state-of-the-art precision for predictions for squark and gluino production at the LHC, we label $\sigma^{\nnllfast}_{h_1 h_2 \to kl}$, which is implemented in the publicly available code package \nnllfast{} which we describe, in the context of the updates for the LHC~Run~3 at $\sqrt{S} = 13.6$~TeV, below in Sec.~\ref{s:nnllfast}.

%%%%%%%%%%
\subsection{Estimation of theoretical uncertainties}\label{s:hocalc:theounc}
%%%%%%%%%%
\paragraph{Factorisation and renormalisation scale uncertainty}
As previously mentioned, we use for our calculations of both the fixed-order as well as the threshold-resummed cross sections a common factorisation and renormalisation scale $\mu$. We vary the scale $\mu$ around a central value chosen as the average mass of the final-state particles, $\mu_0 = \mav$, up and down by a factor of two\footnote{In the calculation of the Coulomb-gluon and bound-state contributions, two additional characteristic scales appear, the Coulomb as well as the Bohr scale, which are different to the common factorisation and renormalisation scale $\mu$, see \cite{Beenakker:2016lwe} for details. When varying $\mu$, we simultaneously vary the Coulomb and Bohr scales up and down by a factor of two. Thus, in the following uncertainties, a variation of these additional scales is implied.}, \mbox{$\mu \in [\mu_0/2, 2\mu_0]$}, to obtain an estimate of the remaining scale dependence and thus missing higher-order corrections. We determine the relative scale uncertainty with respect to the cross section evaluated at the central scale $\sigma_{\mu = \mu_0}$ as:
\begin{equation}\label{eq:scaleunc1}
	\delta_{\mu_0/2} := \frac{\sigma_{\mu = \mu_0/2} - \sigma_{\mu = \mu_0}}{\sigma_{\mu = \mu_0}}\,,\qquad \delta_{2\mu_0} := \frac{\sigma_{\mu = 2\mu_0} - \sigma_{\mu = \mu_0}}{\sigma_{\mu = \mu_0}}\,.
\end{equation}
The lower and upper bounds $\delta_{\mu-}$ and $\delta_{\mu+}$, respectively, of the scale uncertainty are then defined as follows:
\begin{equation}\label{eq:scaleunc2}
	\delta_{\mu+} := \text{max}(\delta_{\mu_0/2}, \delta_{2\mu_0})\,,\qquad  \delta_{\mu-} := \text{min}(\delta_{\mu_0/2}, \delta_{2\mu_0})\,.
\end{equation}
With this definition, the relation $\delta_{\mu-} \le \delta_{\mu+}$ is always ensured. It is possible that both $\delta_{\mu-}$ and $\delta_{\mu+}$ have the same sign, in which case the bounds of the scale uncertainties should be calculated as
\begin{equation}\label{eq:scaleunc3}
	\delta_{\mu+,0} := \text{max}(0, \delta_{\mu+})\,,\qquad  \delta_{\mu-,0} := \text{min}(\delta_{\mu-}, 0)\,,
\end{equation}
so that we always have $\delta_{\mu-,0} \le 0 \le \delta_{\mu+,0}$.

\paragraph{PDF+$\alphas$ uncertainty}
As a choice of parton densities, we make use of the recent PDF4LHC21 set \cite{PDF4LHCWorkingGroup:2022cjn}, which combines the CT18~\cite{Hou:2019efy}, MSHT20~\cite{Bailey:2020ooq}, and NNPDF3.1~\cite{NNPDF:2017mvq} global analyses. The uncertainty associated with the procedure of generating the PDFs is encoded in separate eigenvector or replica sets in the case of a Hessian or a Monte Carlo representation of the set, respectively. Following the prescription in \cite{PDF4LHCWorkingGroup:2022cjn}, they can then be used to determine an additional theoretical uncertainty on the cross sections from the PDF determination.

In our calculations, we use the Hessian set of PDF4LHC21 with one central and 40 eigenvector members as well as $\alphas$ variation, \texttt{PDF4LHC21\_40\_pdfas}, where the requirement of positive-definite PDFs at high $x$ values has been imposed. The relative \mbox{68\% C.\ L.} PDF uncertainty according to the symmetric Hessian prescription is then obtained by computing the cross section for the central, $\sigma^{(0)}$, and each eigenvector set, $\sigma^{(i)}$ with $i = 1,\ldots, N_{\text{set}}$ and $N_{\text{set}} = 40$, as:
\begin{equation}\label{eq:pdfunc}
	\delta_{\text{PDF}} := \frac{1}{\sigma^{(0)}} \sqrt{\sum_{i = 1}^{N_{\text{set}}}\left(\sigma^{(i)} - \sigma^{(0)}\right)^2}
\end{equation}
The PDF4LHC21 set includes additionally two members accounting for the \mbox{68\% C.\ L.} variation of $\alphas$ around its central value $\alphas(M_Z) = 0.118$, corresponding to the lower and upper values $\alphas(M_Z) = 0.117$ and $\alphas(M_Z) = 0.119$, within the determination procedure of the PDFs. We use these members in our calculation to evaluate the $\alphas$ uncertainty associated with the cross section around the central value $\sigma_{\alphas(M_Z) = 0.118}$. The relative $\alphas$ uncertainty is then given as:
\begin{equation}\label{eq:asunc}
    \delta_{\alphas} := \frac{\sigma_{\alphas(M_Z) = 0.119} - \sigma_{\alphas(M_Z) = 0.117}}{2\sigma_{\alphas(M_Z) = 0.118}}\,.
\end{equation}
The combined relative PDF+$\alpha_s$ uncertainty is then obtained as
\begin{equation}\label{eq:pdfasunc}
    \delta_{\text{PDF}+\alphas} := \sqrt{\left(\delta_{\text{PDF}}\right)^2 + \left(\delta_{\alphas}\right)^2}.
\end{equation}

\paragraph{Parametric uncertainty for $\sto_1\sto_1^*$ production}
In the case of stop-antistop production, the cross section depends, in addition to the mass of the produced stops $\mst$, on additional parameters such as the gluino mass $\mgl$, the light-flavoured\footnote{By `light-flavoured' squarks, we mean the superpartners of the light quark flavours.} squark mass $\msq$, as well as the mass of the heavier stop $m_{\sto_2}$ and the stop mixing angle $\theta_\sto$. However, the dependence on these additional parameters is suppressed, since they only appear as loop effects starting from NLO-QCD. We have checked that the dependence of the stop production cross section on $\msq$ and $m_{\sto_2}$ is indeed numerically negligible. We thus fix these values for concreteness to $\msq = 10$~TeV and $m_{\sto_2} = 10.01$~TeV in our computations%
\footnote{The difference of 0.01 TeV between $m_\squ$ and $m_{\sto_2}$ is only of computational nature to avoid numerical divergences in the degenerate case, and the actual choice of values for these masses has a negligible impact on the cross section, see also Table 3 of \cite{Beenakker:2016gmf}.}%
. The effect of a variation of the remaining parameters $\mgl$ and $\theta_\sto$ is in the percent range. In this context, the hierarchy between $\mgl$ and $\mst$ is particularly relevant, since a light gluino facilitates an on-shell decay of the stop into a gluino and a top quark. We therefore keep both $\mst$ and $\mgl$ as variable parameters in our results. Additionally, we encode the effect of varying the stop mixing angle $\theta_\sto$ within the range $\sin(2\theta_\sto) \in [-1, 1]$ in a relative parametric uncertainty $\delta_{\sto,\text{param}\pm}$ with respect to the cross section evaluated with the default value of $\sin(2\theta_\sto) = 0.669$, corresponding to the CMSSM benchmark point 40.2.5 of \cite{AbdusSalam:2011fc}:
\begin{equation}\label{eq:stopparamunc}
	\begin{split}
		\delta_{\sto,\text{param}+} &:= \frac{\operatorname{max}(\sigma_{\sin(2\theta_\sto) \,\in\, [-1, 1]}) - \sigma_{\sin(2\theta_\sto) = 0.669}}{\sigma_{\sin(2\theta_\sto) = 0.669}}\,,\\
		\delta_{\sto,\text{param}-} &:= \frac{\operatorname{min}(\sigma_{\sin(2\theta_\sto) \,\in\, [-1, 1]}) - \sigma_{\sin(2\theta_\sto) = 0.669}}{\sigma_{\sin(2\theta_\sto) = 0.669}}\,.
	\end{split}
\end{equation}

\paragraph{Total theoretical uncertainty}
The total theoretical relative uncertainty $\delta_{\text{tot}\pm}$ on the calculated cross section is then given by all individual uncertainties as discussed above, i.e.\ $\delta_{\mu\pm,0}$ in Eq.~\eqref{eq:scaleunc3} and $\delta_{\text{PDF}+\alphas}$ in Eq.~\eqref{eq:pdfasunc}, added in quadrature,\footnote{A more conservative approach would rely on adding the uncertainties linearly. However, for high enough masses the PDF error vastly dominates, and the difference between adding PDF and scale errors linearly or quadratically is minimal.}
\begin{equation}\label{eq:totunc}
	\delta_{\text{tot}+} := \sqrt{(\delta_{\mu+,0})^2 + (\delta_{\text{PDF}+\alphas})^2}\,,\qquad \delta_{\text{tot}-} := \sqrt{(\delta_{\mu-,0})^2 + (\delta_{\text{PDF}+\alphas})^2}\,,
\end{equation}
including also $\delta_{\sto,\text{param}\pm}$ in Eq.~\eqref{eq:stopparamunc} in the case of stop production:
\begin{equation}\label{eq:stoptotunc}
	\begin{split}
		\delta_{\text{tot}+} := \sqrt{(\delta_{\mu+,0})^2 + (\delta_{\text{PDF}+\alphas})^2 + (\delta_{\sto,\text{param}+})^2}\,,\\
		\delta_{\text{tot}-} := \sqrt{(\delta_{\mu-,0})^2 + (\delta_{\text{PDF}+\alphas})^2 + (\delta_{\sto,\text{param}-})^2}\,.
	\end{split}
\end{equation}
We can then define an upper ($U$) and lower ($L$) limit of the cross section prediction as:
\begin{equation}\label{eq:upperlowerlimits}
	U := \sigma_{\text{central}}\left(1 + \delta_{\text{tot}+}\right),\qquad L := \sigma_{\text{central}}\left(1 - \delta_{\text{tot}-}\right),
\end{equation}
where $\sigma_{\text{central}}$ denotes the cross section calculated with central values for the scale, the PDF member, the $\alphas$ value, and, if applicable, the stop mixing angle. The presented way of treating and combining the uncertainties, now with the modern PDF4LHC21 set, is in agreement with the previous approach taken in \cite{Borschensky:2014cia} for the calculation of squark and gluino cross sections at NLO+NLL accuracy.

%%%%%%%%%%
\subsection{Non-degenerate squark masses}\label{s:hocalc:nondegsquark}
%%%%%%%%%%
As mentioned previously, while calculating the \nnlonnll{} predictions according to Eq.~\eqref{eq:matching}, we assume an 8- or 10-fold degeneracy among the light-flavoured squark masses, so that the cross section only depends on one squark mass parameter $\msq$. To compute the cross section predictions for an MSSM parameter point with non-degenerate squark masses, the parameter $\msq$ should then be chosen as the average value of all light-flavoured squark masses other than $\sto_1$, $\sto_2$ (and $\sbo_1$, $\sbo_2$, if appropriate).

In case cross section predictions for non-degenerate squark masses are needed, we propose as a prescription to rescale the \nnllfast{} cross section obtained by Eq.~\eqref{eq:matching} by the factor
\begin{equation}\label{eq:non-deg-factor}
    R_{\text{non-deg.}} := \frac{\sigma^{\text{LO,\,non-deg.}}_{h_1 h_2 \to kl}(m_{\tilde u_L}, m_{\tilde u_R}, m_{\tilde d_L}, m_{\tilde d_R}, \ldots)}{\sigma^{\text{LO,\,deg.}}_{h_1 h_2 \to kl}(\msq)}\,,
\end{equation}
where $\sigma^{\text{LO,\,non-deg.}}_{h_1 h_2 \to kl}$ is the LO cross section for the squark and gluino production process with all squark masses considered non-degenerate, while $\sigma^{\text{LO,\,deg.}}_{h_1 h_2 \to kl}$ is the corresponding LO cross section with degenerate squark masses, which can both be obtained from e.g.\ \prospino{}. Then, the approximation of the total cross section with non-degenerate squark masses is given as:
\begin{equation}\label{eq:non-deg-nnllfast-xsection}
    \sigma^{\nnllfast{},\,\text{non-deg.}}_{h_1 h_2 \to kl} = R_{\text{non-deg.}} \times \sigma^{\nnllfast}_{h_1 h_2 \to kl}\,.
\end{equation}
Note that this is the same procedure as implemented in the \prospino{} 2 code to compute approximate NLO-QCD predictions for non-degenerate squarks.

The quality of this approximation was studied in the past for selected pre-Run 2 benchmarks points (e.g.\ in \cite{Goncalves-Netto:2012nvl,Gavin:2013kga,Gavin:2014yga}). When a sum of the cross sections over different flavour and chirality combinations was considered, the studies showed only negligible differences between NLO-QCD $K$-factors, i.e.\ ratios of the NLO over the LO cross section, calculated with non-degenerate and degenerate squark masses. Based on the observed behaviour of the NLO cross section, as well as the proportionality of $\Delta$NNLO$_\text{Approx}$ to the LO cross section, we expect similarly negligible effects from non-degenerate squark masses for $\sigma_{h_1 h_2 \to kl}^{\mathrm{NNLO_{Approx}}}$. In addition, the bound-state contributions, as well as the threshold-resummed NNLL corrections are flavour-blind. Therefore the same conclusion must hold, i.e.\ accounting for squark mass degeneracy by rescaling with the ratio of Eq.~\eqref{eq:non-deg-factor} as done in Eq.~\eqref{eq:non-deg-nnllfast-xsection} provides a very good approximation.

%% file: 4-nnllfast.tex
%%%%%%%%%%%%%%%%%%%%%%%%%%%%%%%%%%%%%%%%%%%
\section{\nnllfast}\label{s:nnllfast}
%%%%%%%%%%%%%%%%%%%%%%%%%%%%%%%%%%%%%%%%%%%
The cross sections for squark and gluino hadroproduction at \nnlonnll{} accuracy, evaluated according to Eqs.~\eqref{eq:matching} and~\eqref{eq:nlo-nnloapprox} with the PDF4LHC21 set at the LHC Run 3 collision energy of $\sqrt{S} = 13.6$~TeV, together with all the associated theoretical uncertainties, are provided in the version 2.0 of the publicly available code \nnllfast{}. The package is a successor to the \textsc{NLL-fast} project~\cite{Kulesza:2008jb,Kulesza:2009kq,Beenakker:2009ha,Beenakker:2010nq,Beenakker:2011fu,Beenakker:2011dk,Borschensky:2014cia,Beenakker:2015rna}.

The \nnllfast{} code consists of pre-computed total cross sections and uncertainties provided as numerical grids, together with a fast interpolation code. All processes described in Sec.~\ref{s:squgluprod} are implemented. In addition to these processes, we also provide predictions for gluino-pair production in the limit of decoupled, i.e.\ very heavy, squarks, and squark-antisquark production in the limit of decoupled gluinos. The mass ranges of $m_{\squ/\sto_1}$ and $m_\glu$ for the grids are the following:\footnote{Extended mass ranges are available on request. For specific processes, tabulated cross sections for mass values outside of the mentioned ranges are available on the TWiki page of the LHC SUSY Cross Section Working Group \url{https://twiki.cern.ch/twiki/bin/view/LHCPhysics/SUSYCrossSections}.}
\begin{itemize}
	\item $\glu\glu$, $\squ\squ^*$, $\squ\glu$, and $\squ\squ$ production:
		\begin{equation}\label{eq:massrangeggsbsgss}
			\msq \in [500, 3000] \text{ GeV},\qquad \mgl \in [500, 3000] \text{ GeV},
		\end{equation}
	\item $\glu\glu$ production with decoupled squarks ($\msq$ chosen very heavy):
		\begin{equation}\label{eq:massrangegdcpl}
			m_\glu \in [500, 3000] \text{ GeV},
		\end{equation}
	\item $\squ\squ^*$ production with decoupled gluinos ($\mgl$ chosen very heavy):
		\begin{equation}\label{eq:massrangesdcpl}
			\msq \in [500, 3000] \text{ GeV},
		\end{equation}
	\item $\sto_1\sto_1^*$ production:
		\begin{equation}\label{eq:massrangest}
			\mst \in [100, 3000] \text{ GeV},\qquad \mgl \in [500, 5000] \text{ GeV}.
		\end{equation}
\end{itemize}
For the computation of the grid points at \nnlonnll{} accuracy, we employ the following codes. The NLO SUSY-QCD cross section is computed using the \prospino{}~2 code \cite{Beenakker:1996ed}. The remaining terms in Eqs.~\eqref{eq:matching} and~\eqref{eq:nlo-nnloapprox}, i.e.\ the threshold-enhanced approximated NNLO corrections $\Delta\sigma^{\text{NNLO}_{\text{Approx}}}$, the bound-state contributions $\sigma^{\text{BS}}$, as well as the soft-gluon and Coulomb resummed contributions beyond NNLO accuracy are calculated and cross-checked with two in-house codes, for which we find very good numerical agreement. The uncertainties are computed according to Eqs.~\eqref{eq:scaleunc2} and \eqref{eq:pdfasunc} for the variation of the renormalisation and factorisation scales and the combined PDF+$\alphas$ uncertainty, respectively, and, in the case of stop-antistop production, according to Eq.~\eqref{eq:stopparamunc} for the variation of the remaining SUSY parameters.

Compared to the previous version 1.1 of \nnllfast{}, the technical work on the version 2.0 update consisted of generating the new grids containing the NLO-QCD and the threshold-resummation-improved \nnlonnll{} cross sections together with the associated uncertainties, evaluated with the PDF4LHC21 set at $\sqrt{S} = 13.6$~TeV. Additionally, checks regarding the interpolation quality for cross sections in between grid points were performed, to make sure that the interpolated results are in agreement with those obtained from a direct computation. We found up to 2\% discrepancies in between grid points for processes other than $\sto_1\sto_1^*$ production, where the interpolation error could reach 5\%. These maximal values were encountered mostly at the extreme edges of the grids (i.e.\ very small or very large masses). For $\glu\glu$ production, interpolation errors of 1--2\% are observed also for some intermediate points. Other than these singular cases, the interpolation accuracy was found to always be better than 1\%.

%%%%%%%%%%
\subsection{Running of the code}\label{s:nnllfast:runningcode}
%%%%%%%%%%
The \nnllfast{} 2.0 code and its previous versions are made available under the following link:
\begin{center}
	\url{https://www.uni-muenster.de/Physik.TP/~akule_01/nnllfast}
\end{center}
After downloading the \nnllfast{} 2.0 package and unpacking it, the interpolation code written in Fortran can be compiled within the \texttt{nnllfast-2.0/} directory by typing in a terminal the following command, assuming the GNU Fortran compiler of the GNU Compiler Collection to be used:
\begin{lstlisting}
	gfortran nnllfast-2.0.f -o name_of_the_executable
\end{lstlisting}
\begin{table}[tp]
	\centering
	\begin{tabular}{cc}
		\toprule
		Label & Production process \\
		\midrule
		\verb+gg+ & $\glu\glu$ \\
		\verb+sb+ & $\squ\squ^*$ \\
		\verb+sg+ & $\squ\glu$ \\
		\verb+ss+ & $\squ\squ$ \\
		\verb+st+ & $\sto_1\sto_1^*$ \\
		\verb+gdcpl+ & $\glu\glu$ with decoupled squarks \\
		\verb+sdcpl+ & $\squ\squ^*$ with decoupled gluinos \\
		\bottomrule
	\end{tabular}
	\caption{Abbreviations for the production processes of squark, gluino, and stop production when running the \nnllfast{} code. For specific running signatures, in particular for the \texttt{st}, \texttt{gdcpl}, and \texttt{sdcpl} processes, see the corresponding text below.}
	\label{tab:processes}
\end{table}
The name of the executable \verb+name_of_the_executable+ can be chosen freely. The executable can then be called to obtain cross section results including the associated theoretical uncertainties:
\begin{lstlisting}
	./name_of_the_executable <process> <squark_mass> <gluino_mass>
\end{lstlisting}
where \verb+<process>+ is one of the labels listed in Table~\ref{tab:processes}, and \verb+<squark_mass>+ as well as \verb+<gluino_mass>+ correspond to the pair of values for $\msq$ as well as $\mgl$ for which the cross section should be output. In the case of stop production, the second argument of the squark mass \verb+<squark_mass>+ is replaced by \verb+<stop_mass>+ corresponding to the light stop mass $\mst$:
\begin{lstlisting}
	./name_of_the_executable st <stop_mass> <gluino_mass>
\end{lstlisting}
In the case of gluino-pair production with decoupled squarks or squark-antisquark production with decoupled gluinos, the executable should be called as:
\begin{lstlisting}
	./name_of_the_executable gdcpl <gluino_mass>
\end{lstlisting}
with \verb+<gluino_mass>+ set to the value of choice for $\mgl$, or
\begin{lstlisting}
	./name_of_the_executable sdcpl <squark_mass>
\end{lstlisting}
with \verb+<squark_mass>+ set to the required value of $\msq$, respectively.

An example output for the following command line
\begin{lstlisting}
	./name_of_the_executable sg 1700 2100
\end{lstlisting}
is:
\begin{lstlisting}[basicstyle=\ttfamily\scriptsize]
# LHC @ 13.6 TeV, NNLO PDF4LHC21 (LHAPDF ID 93300)
# process: sg
# ms[GeV] mg[GeV]  NLO[pb]  NNLL+NNLO_app[pb] d_mu+[%] d_mu-[%] d_pdfas+[%] d_pdfas-[%] K_NNLL
-----------------------------------------------------------------------------------------------
  1700.   2100.   0.733E-02    0.891E-02        3.69    -5.49     9.86       -9.86       1.21
\end{lstlisting}
Here, the first two columns denote the input values of $\msq$ and $\mgl$, the third column corresponds to the fixed-order NLO-QCD cross section, and the fourth column corresponds to the \nnllfast{} cross section of Eq.~\eqref{eq:matching} at \nnlonnll{} accuracy including threshold-resummation corrections. Columns five to eight correspond to the upper and lower scale uncertainties $\delta_{\mu\pm}$ of Eq.~\eqref{eq:scaleunc2} and the PDF+$\alphas$ uncertainty $\delta_{\text{PDF}+\alphas}$ of Eq.~\eqref{eq:pdfasunc} given in percent. In the last column, we output the $K_{\text{NNLL}}$ factor given as the ratio of the \nnllfast{} cross section over the NLO-QCD result,
\begin{equation}\label{eq:knnllfactor}
	K_{\text{NNLL}} := \frac{\sigma^{\nnllfast}}{\sigma^{\text{NLO}}}\,,
\end{equation}
which denotes the size of the threshold-enhanced corrections beyond NLO.

%% file: 5-numresults.tex
%%%%%%%%%%%%%%%%%%%%%%%%%%%%%%%%%%%%%%%%%
\section{Numerical results}\label{s:numerics}
%%%%%%%%%%%%%%%%%%%%%%%%%%%%%%%%%%%%%%%%%
In this section, we present numerical results based on \nnlonnll{} calculations which can be obtained with the \nnllfast{} package. Note that here and in the following, we use for the accuracy the labels ``\nnlonnll{}'' and ``\nnllfast{}'' interchangeably, and we always mean our best accuracy by including all terms according to Eqs.~\eqref{eq:matching}. Unless otherwise stated, the results are obtained with the PDF4LHC21 Hessian set (LHAPDF ID 93300), accessed through the \textsc{LHAPDF}~6 library \cite{Buckley:2014ana}, and at a centre-of-mass energy of $\sqrt S=13.6$~TeV, using a common renormalisation and factorisation scale $\mu$ which has been set to the central scale choice of the average mass of the produced particles, $\mu = \mu_0 = \mav$, as discussed above. We note that the PDF4LHC collaboration offers their recent PDF4LHC21 sets only at NNLO accuracy, so that in the following discussion, both the NLO as well as the \nnlonnll{} cross sections are evaluated with the same NNLO PDFs.

As an additional remark, we note that our calculation including threshold resummation would in principle require using a threshold-improved PDF set, such as the one from \cite{Bonvini:2015ira}, for numerical predictions. However, as there is no threshold-improved PDF set based on more recent data, we consider using modern sets such as PDF4LHC21, albeit determined on the basis of only fixed-order predictions, preferable. In \cite{Beenakker:2015rna}, the effect of the PDF set from \cite{Bonvini:2015ira} on squark and gluino production was studied at NLO+NLL accuracy, and it was found that the difference between a conventional fixed-order PDF set and the one including threshold resummation effects is contained within the total PDF uncertainty of the conventional set.

%%%%%%%%%%
\subsection{Predictions for $\sqrt{S} = 13.6$ TeV}\label{s:numerics:pred136}
%%%%%%%%%%
\begin{figure}[tp]
	\centering
	\includegraphics[width=0.5\textwidth]{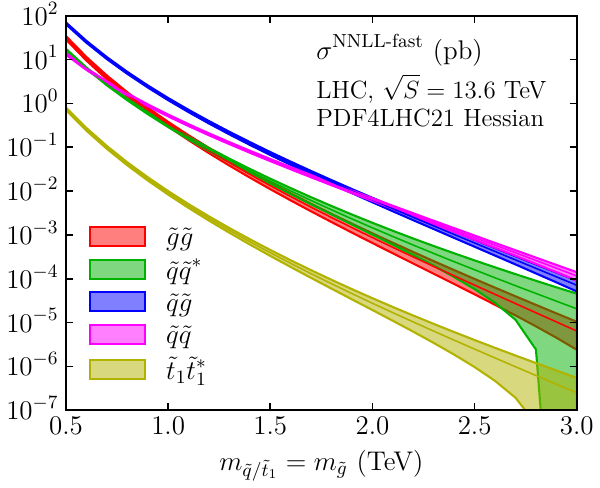}\includegraphics[width=0.5\textwidth]{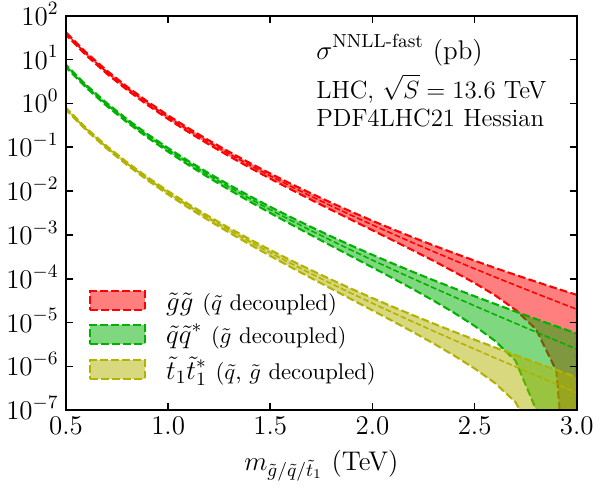}
	\caption{Total inclusive cross sections for squark, gluino, and stop production at the LHC at $\sqrt{S} = 13.6$~TeV. Left: equal $\msq = \mgl$ (or $\mst = \mgl$ for $\sto_1\sto_1^*$, respectively). Right: decoupled scenarios. The bands denote the uncertainty envelope around the central cross section values as discussed in Eq.~\eqref{eq:upperlowerlimits}. As we consider all light-flavour squarks to be degenerate, the letter $\squ$ implicitly stands for the sum over all squark flavours, and the production cross sections consist of the sum of each possible flavour combination.}
	\label{fig:nnllfast20}
\end{figure}
The dependence of the total cross section for all processes of interest on the mass of the produced sparticle in the final state is shown in Fig.~\ref{fig:nnllfast20}. The central cross section values as well as the uncertainty bands are computed according to Eq.~\eqref{eq:upperlowerlimits}. The width of the bands denotes the total theoretical uncertainty, calculated as described in section~\ref{s:hocalc:theounc}. The left plot of Fig.~\ref{fig:nnllfast20} displays the cross sections under the assumption of equal squark and gluino masses, $\msq = \mgl$ (or $\mst = \mgl$ for stop-antistop production), while the right plot presents the case of the decoupled scenarios, i.e.\ where either the squarks (in the case of $\glu\glu$), or the gluinos ($\squ\squ^*$), or both ($\sto_1\sto_1^*$) are assumed very heavy\footnote{We checked that for $\sto_1\sto_1^*$, while the squarks are always chosen to be decoupled at $\msq = 10$~TeV due to their negligible impact on the cross section, a value of $\mgl = 5$~TeV is sufficiently high to consider stop production in the decoupling regime, as the cross section remains constant even for higher gluino masses such as $\mgl = 10$~TeV.}. In the cases of $\squ\squ^*$, $\squ\glu$, and $\squ\squ$, the ten light-flavour squarks are considered as degenerate, i.e.\ they all have the same mass $\msq$, and the production cross section shown corresponds to a sum over all degenerate final states.

For equal squark and gluino masses, it can be seen in the left plot of Fig.~\ref{fig:nnllfast20} that while for low masses, the processes where one or two gluinos are being produced dominate over the other processes, the cross section of squark-pair production drops less rapidly and becomes the dominant process for large squark and gluino masses. This effect is related to the parton luminosities: the process $\squ\squ$ can proceed via the collision of two valence quarks, while all other processes depend, at LO, through their initial states on antiquark or gluon PDFs, which, at high masses and consequently high momentum fractions of the partons, are more strongly suppressed than the valence-quark PDFs. In the case of the decoupled scenarios, the right plot of Fig.~\ref{fig:nnllfast20} shows that, while the $\glu\glu$ cross section is of similar size for small masses as in the equal-mass case, it reaches larger values for large gluino masses as compared to the equal-mass case. The opposite behaviour can be seen for $\squ\squ^*$, where the cross section of the decoupled case always lies below the equal-mass case. In the case of $\sto_1\sto_1^*$, there is almost no difference between the equal-mass and decoupled cases due to the dependence on $\msq$ and $\mgl$ arising only from higher orders.

There exists no decoupling limit for $\squ\glu$ as both squarks and gluinos appear in the final state, and the cross section thus tends towards zero for very heavy $\msq$ or $\mgl$. Similarly, the $\squ\squ$ cross section becomes zero for decoupled $\glu$, as can be seen e.g.\ from the tree-level diagrams of Fig.~\ref{fig:losqglproduction} (d), where the gluino appears as a virtual particle in all diagrams in the $t$- or $u$-channel, respectively, and the amplitudes are thus heavily suppressed for very large $\mgl$.

We note that for $\squ\squ^*$ and $\sto\sto^*$ in the equal-mass case as well as for all processes in the decoupled case, the uncertainty band towards large mass values becomes very large and the error surpasses 100\%, causing the lower end of the band to extend towards very small values in the plots with a logarithmic axis.

%%%%%%%%%%
\subsubsection{PDF+$\alphas$ and scale uncertanties}\label{s:numerics:pred136:unc}
%%%%%%%%%%
\begin{figure}[tp]
	\centering
	\includegraphics[width=0.5\textwidth]{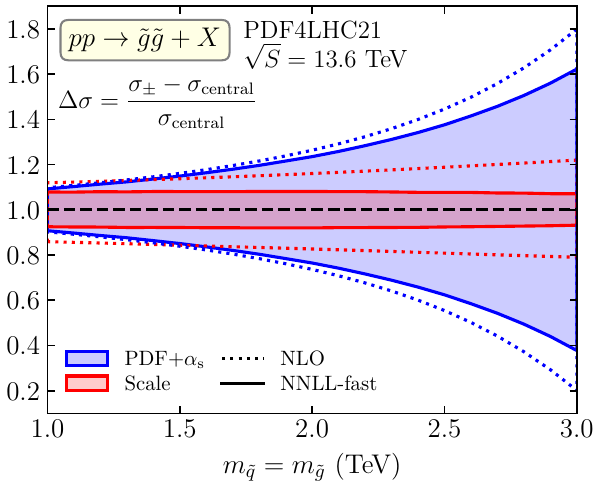}\includegraphics[width=0.5\textwidth]{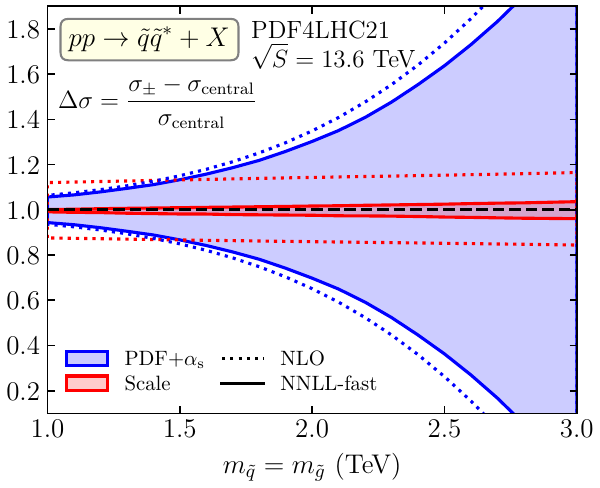}\\
	\includegraphics[width=0.5\textwidth]{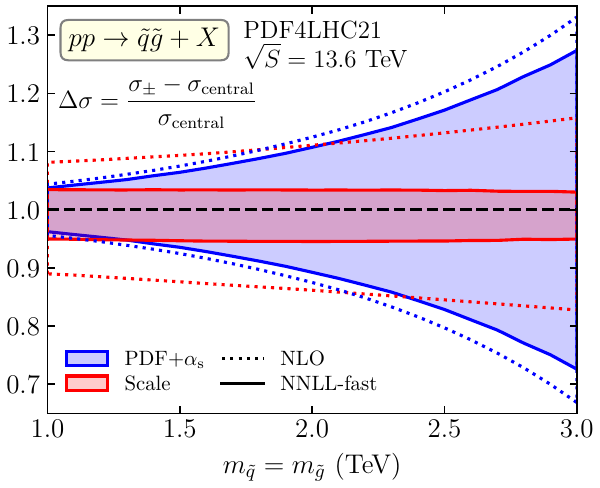}\includegraphics[width=0.5\textwidth]{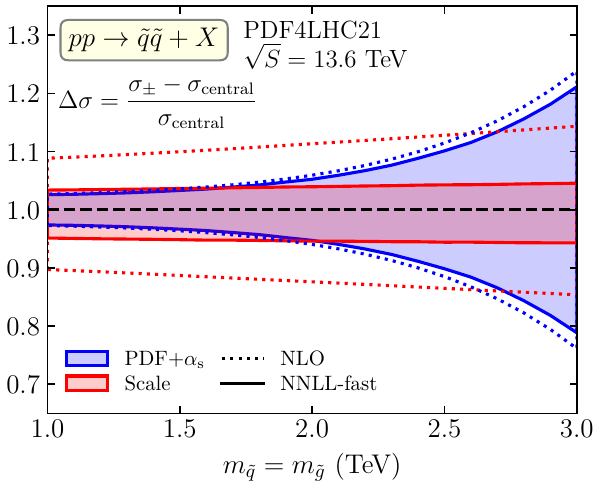}
	\caption{Theoretical uncertainties for $\glu\glu$, $\squ\squ^*$, $\squ\glu$, and $\squ\squ$ production in the case of equal $\msq = \mgl$ at \nnlonnll{} with a centre-of-mass energy of $\sqrt{S} = 13.6$~TeV and the PDF4LHC21 set, as provided by \nnllfast{}~2.0. For the scale uncertainties, $\sigma_{\pm} := \sigma_{\text{central}}\left(1 \pm \delta_{\mu\pm,0}\right)$ with $\delta_{\mu\pm,0}$ from Eq.~\eqref{eq:scaleunc3}, and for the PDF+$\alphas$ uncertainties, $\sigma_{\pm} := \sigma_{\text{central}}\left(1 \pm \delta_{\text{PDF}+\alphas}\right)$ with $\delta_{\text{PDF}+\alphas}$ from Eq.~\eqref{eq:pdfasunc}.}
	\label{fig:uncertaintiesnew}
\end{figure}
\begin{figure}[tp]
	\centering
	\includegraphics[width=0.5\textwidth]{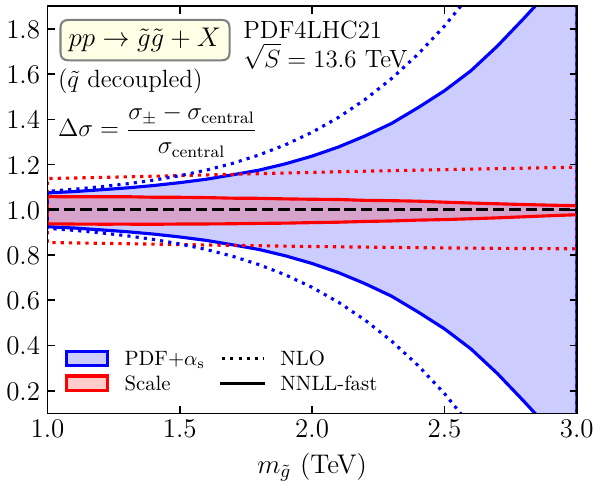}\includegraphics[width=0.5\textwidth]{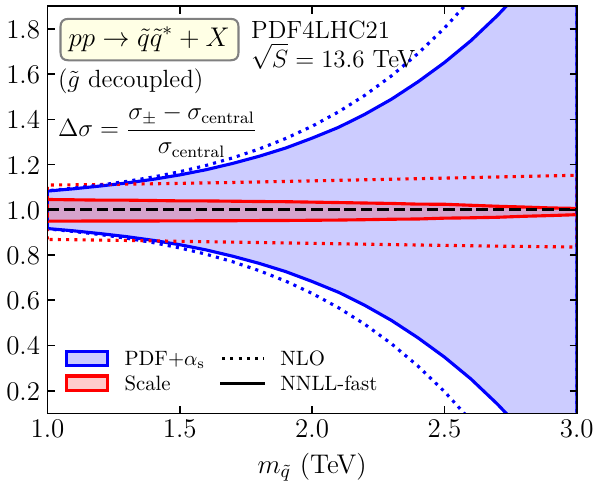}\\
	\includegraphics[width=0.5\textwidth]{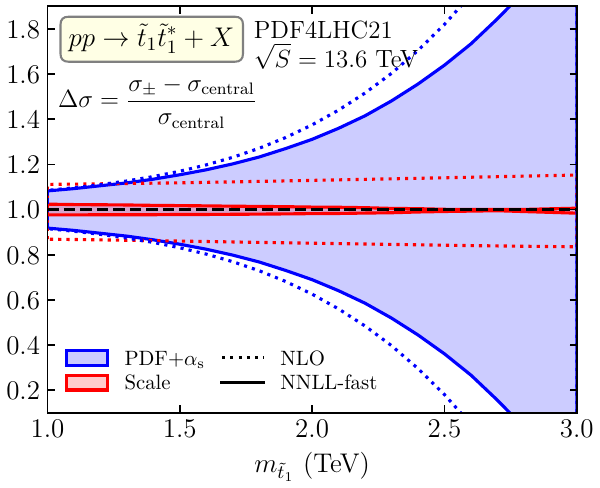}
	\caption{Theoretical uncertainties for $\glu\glu$, $\squ\squ^*$, and $\sto_1\sto_1^*$ production in the decoupled scenarios, where all sparticles other than the produced ones are assumed very heavy, with the otherwise same set-up as for Fig.~\ref{fig:uncertaintiesnew}.}
	\label{fig:uncertaintiesnewdec}
\end{figure}
In the following, we discuss the sources of theoretical uncertainties for the processes of squark, gluino, and stop production, as discussed in Sec.~\ref{s:hocalc:theounc}. In Figs.~\ref{fig:uncertaintiesnew} (for the equal-mass case) and \ref{fig:uncertaintiesnewdec} (for the decoupled scenarios), we show the relative sizes of the PDF+$\alphas$ uncertainties as well as the uncertainty related to the variation of the common renormalisation and factorisation scale $\mu$.

We find that going from the best fixed-order prediction NLO to \nnlonnll{}, the theoretical scale uncertainty is reduced significantly for all processes and is almost constant with respect to the masses of the produced particles. The strongest reduction is found for $\squ\squ^*$ in the equal-mass case, and for $\sto_1\sto_1^*$ in the decoupled scenario. For the shown mass ranges, the scale uncertainties are of the order of or below 10\% for all processes.

In contrast, the uncertainty due to the parametrisation of the PDFs as well as a variation of the value of $\alphas$ is not affected by the increase in accuracy to the same degree as the scale uncertainty. As noted before, both the NLO as well as \nnlonnll{} cross sections are computed with the same PDFs at NNLO accuracy, so we do not expect a significant improvement of the PDF uncertainties. We nonetheless see a slight decrease of the PDF+$\alphas$ uncertainty, in particular towards higher masses where the uncertainty becomes large, which is related to cancellations of higher-order terms between the PDF evolution and the threshold effects beyond NLO. After including resummation corrections, the PDF+$\alphas$ uncertainty now constitutes the dominant source of uncertainty for all processes, with the exception of $\squ\squ$ where the scale uncertainty is of the same order or slighly above the PDF+$\alphas$ uncertainty up to about $\msq = \mgl = 2$~TeV. As mentioned already in the discussion of Fig.~\ref{fig:nnllfast20}, for the processes of $\squ\squ^*$ and $\sto_1\sto_1^*$ as well as $\glu\glu$ in the decoupled case, the PDF+$\alphas$ uncertainty grows above 100\% at high masses, where the gluon initial states dominate, due to a lack of data to constrain in particular the gluon and sea quark luminosities at high scales.

%%%%%%%%%%
\subsection{Comparison to previous results for $\sqrt{S} = 13$ TeV}\label{s:numerics:compare13}
%%%%%%%%%%
\begin{figure}[tp]
	\centering
	\includegraphics[width=\textwidth]{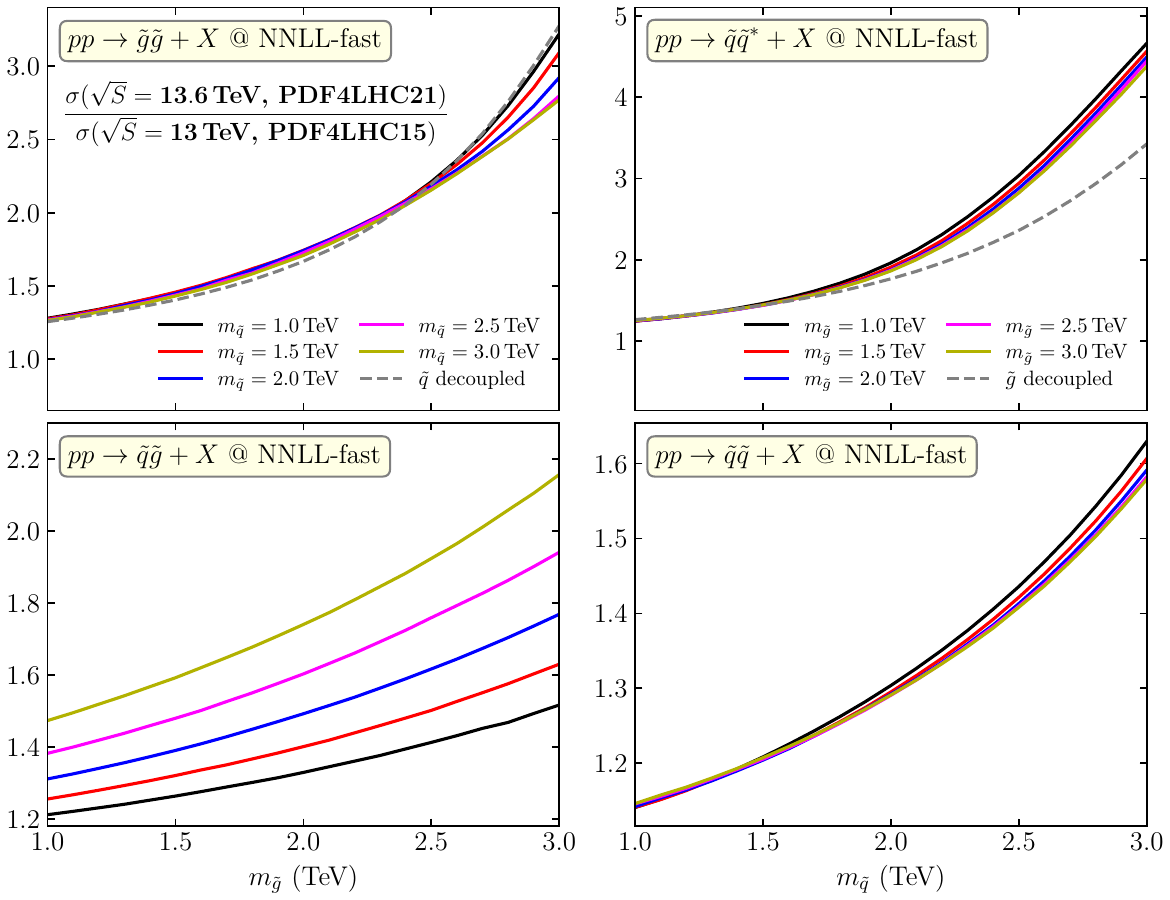}\\
	\includegraphics[width=0.5\textwidth]{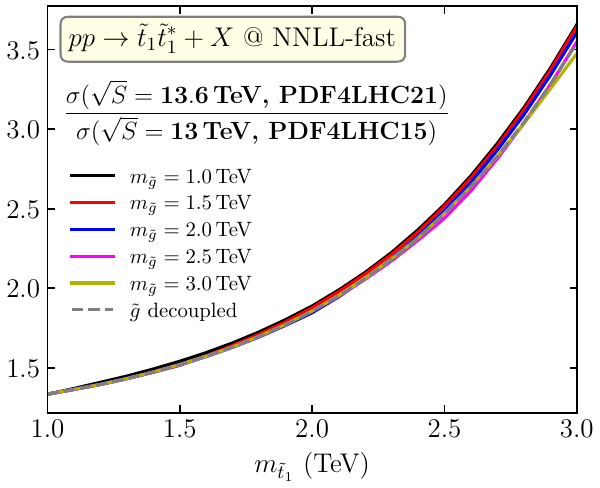}
	\caption{Comparison between the central \nnlonnll{} cross sections for squark, gluino, and stop production as provided by \nnllfast{}~2.0 (for $\sqrt{S} = 13.6$~TeV and with PDF4LHC21) and the previous version \nnllfast{}~1.1 (for $\sqrt{S} = 13$~TeV and with PDF4LHC15), presented as the ratio of the two. The ratios are shown depending on the mass of the produced sparticle for a range of values of the other mass parameter.}
	\label{fig:nnllfast20vs11SandPDF}
\end{figure}
We move on to a discussion of the differences between our current results for Run 3 as presented in this paper and the previous predictions from 2016 for Run 2 at $\sqrt{S} = 13$~TeV \cite{Beenakker:2016lwe} and computed with the PDF4LHC15 set \cite{Butterworth:2015oua}. The comparison is shown in Fig.~\ref{fig:nnllfast20vs11SandPDF} as a ratio of the central \nnlonnll{} cross sections obtained by the recent \nnllfast{}~2.0 to the previous \nnllfast{}~1.1 results for each process. We probe the parameter space by presenting the dependence on the mass of the produced sparticle for a selected range of values of the other mass parameter ($\mgl$ in the case of squark production, $\msq$ for the gluino production). The decoupled cases, as discussed before, are denoted in the plots by the dashed lines, wherever applicable. For all processes, the ratio is growing with increasing masses of the produced particles. With the exception of $\squ\glu$ production, the dependence of the ratio on the other mass parameter is relatively small, and in most cases begins to be visible only for very large masses of the produced particles. The modification of the cross section, illustrated by the ratio, ranges from a few tens of percent to a factor of a few, with the highest factor of about 4.5 observed for $\squ\squ^*$, and the smallest of about 1.6 for $\squ\squ$ production at high masses.

\begin{figure}[tp]
	\centering
	\includegraphics[width=\textwidth]{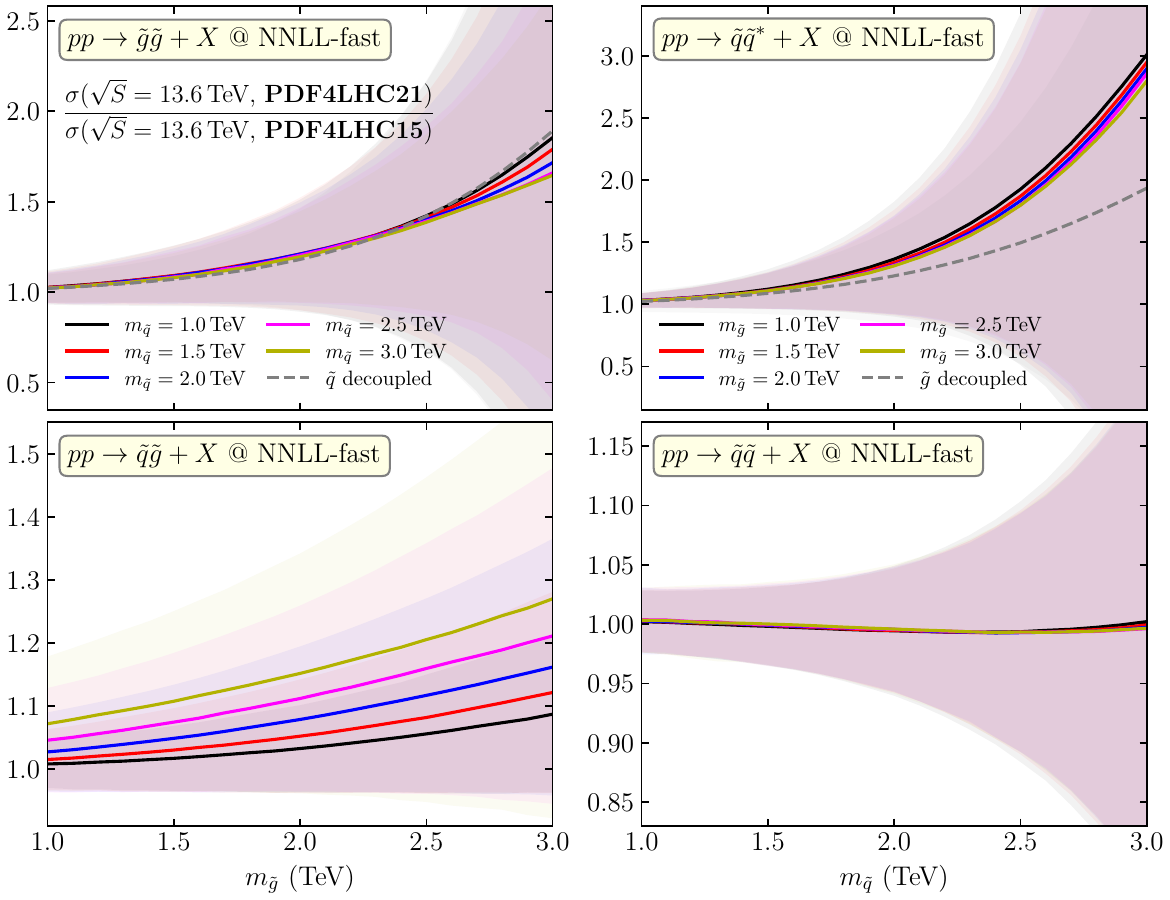}\\
	\includegraphics[width=0.5\textwidth]{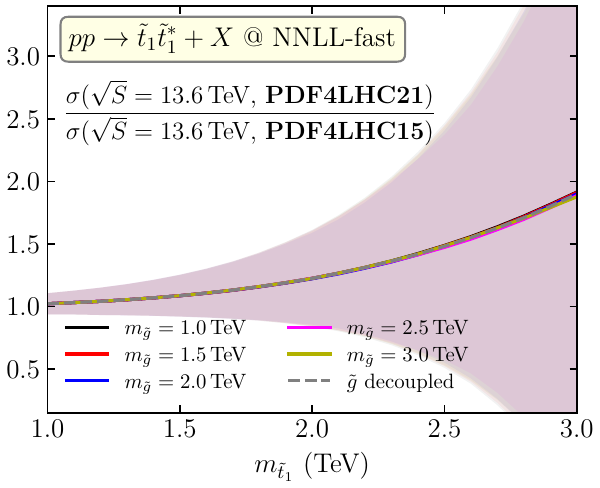}
	\caption{Same as Fig.~\ref{fig:nnllfast20vs11SandPDF}, but with a fixed centre-of-mass energy ($\sqrt{S} = 13.6$~TeV) and different PDFs used in the numerator (PDF4LHC21) and denominator (PDF4LHC15). The shaded bands denote the size of the PDF+$\alphas$ uncertainties from the PDF4LHC21 set around the central cross section predictions in the numerator, in the same colour scheme as the lines.}
	\label{fig:nnllfast20vs11PDF}
\end{figure}
\begin{figure}[tp]
	\centering
	\includegraphics[width=\textwidth]{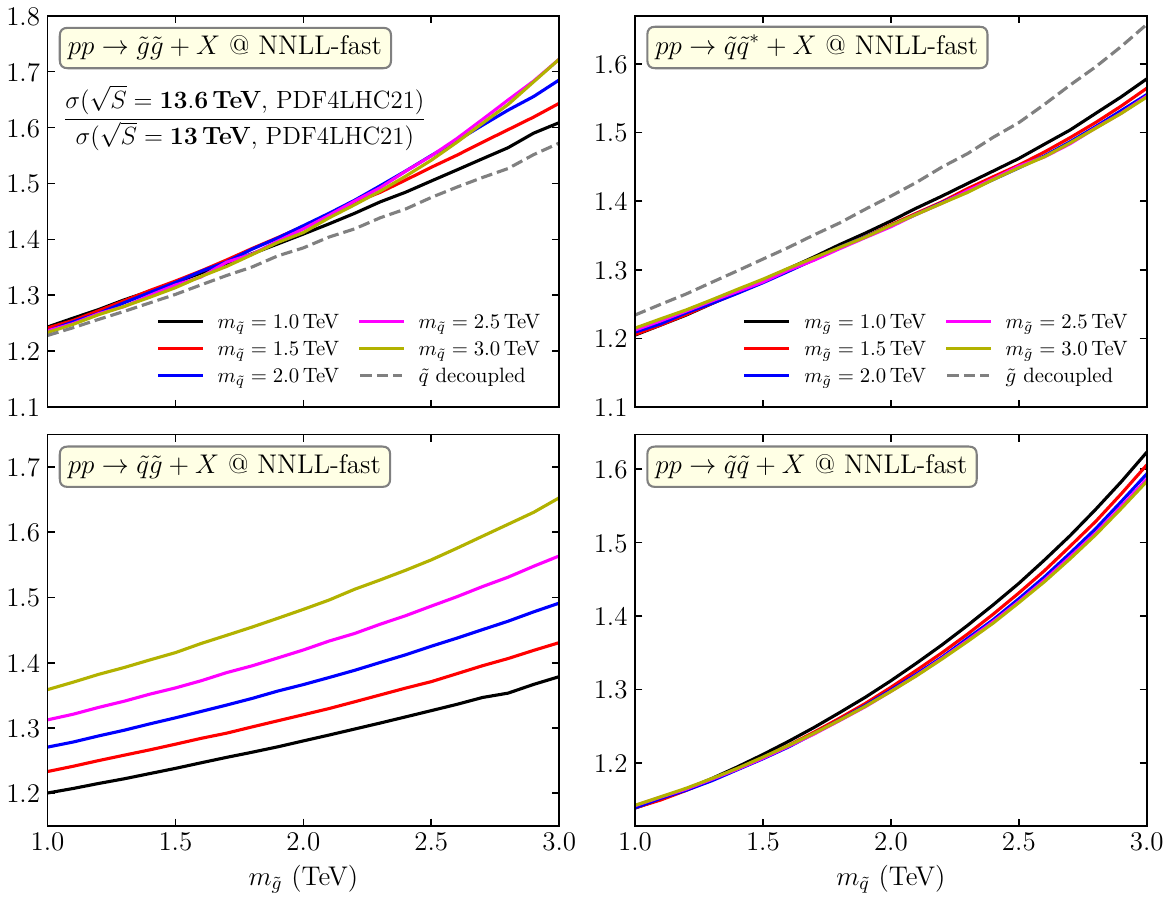}\\
	\includegraphics[width=0.5\textwidth]{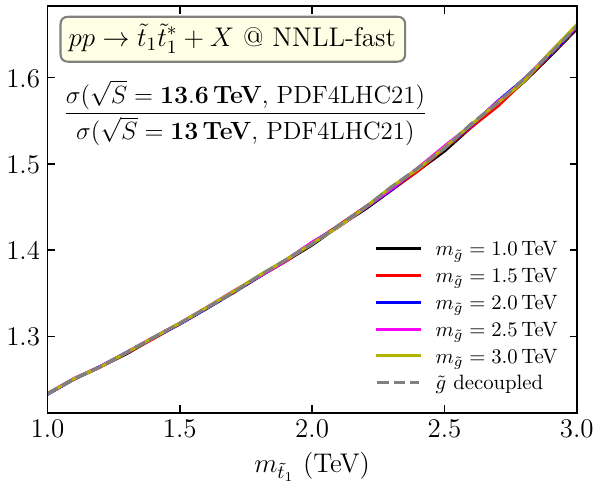}
	\caption{Same as Fig.~\ref{fig:nnllfast20vs11SandPDF}, but with a fixed PDF set (PDF4LHC21) and different centre-of-mass energies used in the numerator ($\sqrt{S} = 13.6$~TeV) and the denominator ($\sqrt{S} = 13$~TeV).}
	\label{fig:nnllfast20vs11S}
\end{figure}
It is interesting to study where the effect is coming from. To this end, in Figs.~\ref{fig:nnllfast20vs11PDF} and~\ref{fig:nnllfast20vs11S}, we show separately the impact of the change of the PDF set and of the increase in the collision energy, respectively. While the ratio for changing the PDF set is strongly influenced by the different processes depending on different PDF luminosities, the increase due to changing the centre-of-mass energy is approximately the same for all processes and accounts for up to about 60--70\% at high masses. In particular, we find that the increase in the $\squ\squ$ cross section is solely driven by calculating it at a higher centre-of-mass energy, whereas for all other processes the increase can be traced back to both higher energy collisions and the different set of PDFs. This is in agreement with the $\squ\squ$ production taking place only in the $qq$ channel at LO and the $qq$ luminosities being very similar for both PDF sets~\cite{PDF4LHCWorkingGroup:2022cjn}. Additionally, we show in Fig.~\ref{fig:nnllfast20vs11PDF} the size of the PDF+$\alphas$ uncertainties plotted around the central cross section predictions in the numerator, which are calculated with the PDF4LHC21 set. While the effect of updating the PDF set from PDF4LHC15 to PDF4LHC21 is, for the shown mass range, contained entirely within the uncertainty bands, the uncertainties grow in particular for the high-mass region to very large values, which highlights the need for a precise determination of PDFs in the large-$x$ region in order to properly constrain the squark and gluino processes at large $m_\squ$ and $m_\glu$.

\begin{figure}[tp]
	\centering
	\includegraphics[width=\textwidth]{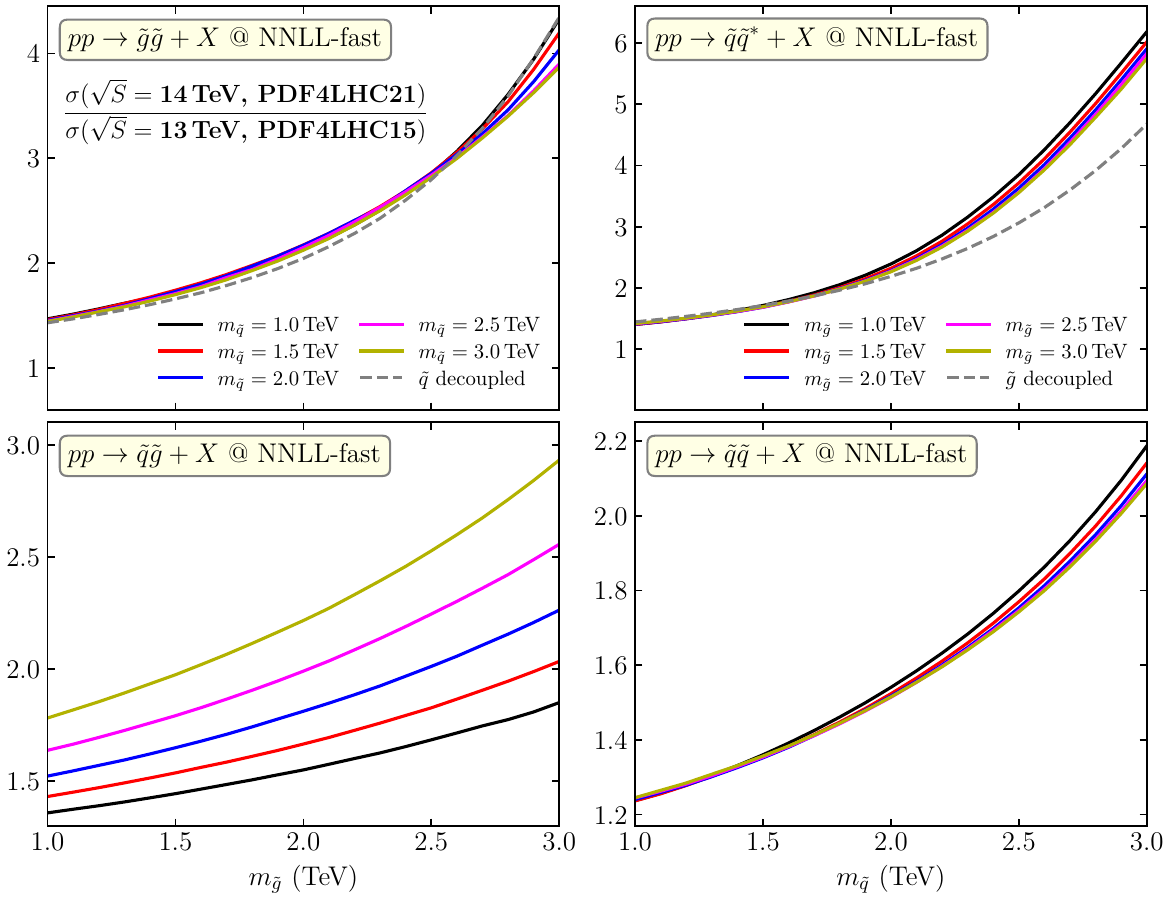}\\
	\includegraphics[width=0.5\textwidth]{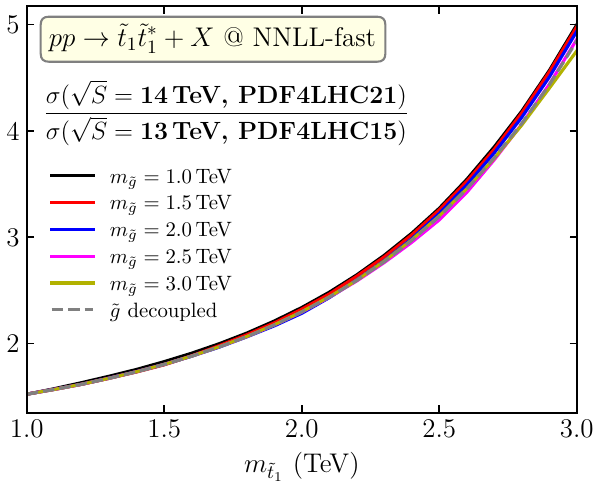}
	\caption{Same as Fig.~\ref{fig:nnllfast20vs11SandPDF}, but with an increased centre-of-mass energy of $\sqrt{S} = 14$~TeV in the numerator.}
	\label{fig:14vsold}
\end{figure}
An analogous comparison of the \nnlonnll{} results obtained for a centre-of-mass energy of $\sqrt S = 14$~TeV with the PDF4LHC21 set and the results for Run 2 provided by \nnllfast{}~1.1 is presented in Fig.~\ref{fig:14vsold} and shows higher values of the corresponding ratios, but a qualitatively similar dependence on the $\squ$ and $\glu$ masses as in the case of $\sqrt S = 13.6$~TeV in Fig.~\ref{fig:nnllfast20vs11SandPDF}.

%%%%%%%%%%
\subsection{Comparison between $K$-factors for $\sqrt{S} = \{13, 13.6, 14\}$ TeV}\label{s:numerics:kfactors}
%%%%%%%%%%
\begin{figure}[tp]
	\centering
	\includegraphics[width=0.5\textwidth]{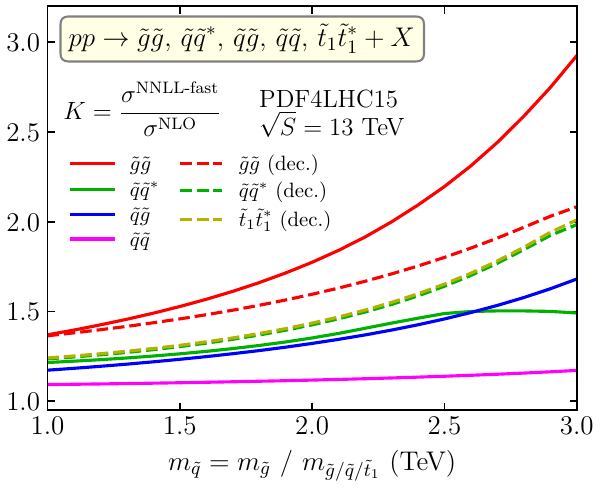}\includegraphics[width=0.5\textwidth]{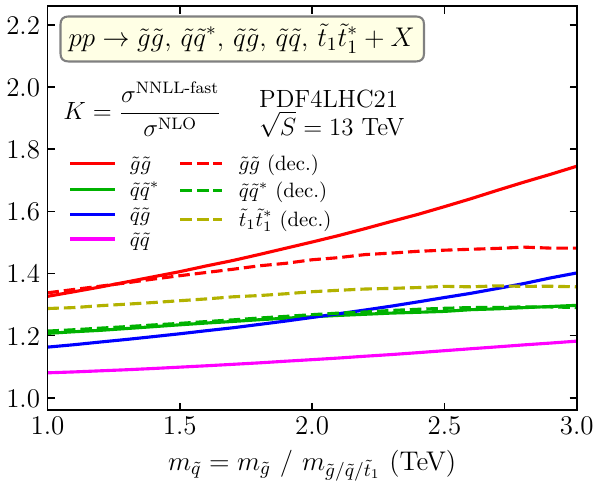}\\
	\includegraphics[width=0.5\textwidth]{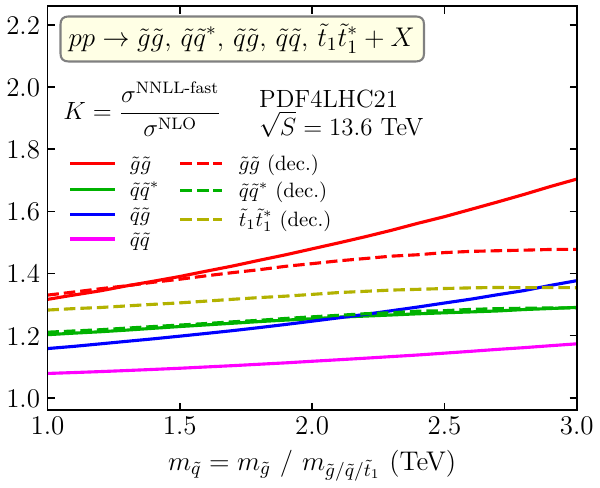}\includegraphics[width=0.5\textwidth]{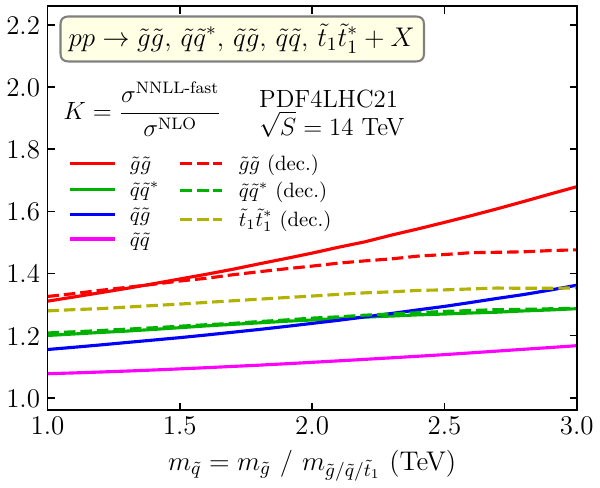}
	\caption{$K$-factors for all processes of squark and gluino production as well as stop production in the decoupled scenario. Top: $\sqrt{S} = 13$~TeV with PDF4LHC15 (left) and PDF4LHC21 sets (right). Bottom: $\sqrt{S} = 13.6$~TeV (left) and $\sqrt{S} = 14$~TeV (right), both with the PDF4LHC21 set. Whenever NLO PDFs are available, i.e.\ in the case of the PDF4LHC15 set, we use NLO PDFs for the calculation of $\sigma^{\text{NLO}}$ in the denominator. In all other cases, we use NNLO PDFs.}
	\label{fig:kfactor}
\end{figure}
Another important information in the context of \nnlonnll{} calculations is the size of the NNLL corrections, as compared to the best fixed-order predictions, i.e.\ NLO. In Fig.~\ref{fig:kfactor}, we thus show the corresponding $K$-factor as defined in Eq.~\eqref{eq:knnllfactor}, i.e.\ the ratio between the central \nnlonnll{} cross sections to the NLO results, for four different set-ups:
\begin{itemize}
	\item for $\sqrt S = 13$~TeV with the PDF4LHC15 set, as provided by \nnllfast{}~1.1,
	\item for $\sqrt S = 13$~TeV with the PDF4LHC21 set,
	\item for $\sqrt S = 13.6$~TeV with the PDF4LHC21 set, as provided by \nnllfast{}~2.0,
	\item and for $\sqrt S = 14$~TeV with the PDF4LHC21 set.
\end{itemize}
Since the PDF4LHC21 set contains only the NNLO PDFs, we use them to also calculate the NLO cross sections entering the $K$-factors, contrary to the case of the PDF4LHC15 set where we use the NLO PDFs to compute the NLO cross sections in the $K$-factors. As expected, we observe that the relevance of resummed corrections diminishes slightly as the collision energy grows bigger. This can be deduced by comparing the $K$-factors for $\sqrt S = 13.6$~TeV and $\sqrt S = 14$~TeV, which are calculated with the same set of PDFs. Comparing the two $K$-factors obtained for $\sqrt S = 13$~TeV with the PDF4LHC15 and PDF4LHC21 sets shows however, that also the results for the $K$-factors are influenced by the PDFs, and that in a much stronger way than by the value of the collision energy. It should be noted that while all $K$-factors computed with the same NNLO PDF4LHC21 set behave similarly for different values of $\sqrt S$, the qualitative differences for the $K$-factors as provided by \nnllfast{}~1.1 are mainly due to NNLO PDFs of PDF4LHC15 being used for the \nnlonnll{} cross sections in the numerator, and NLO PDFs of PDF4LHC15 being used for the NLO cross sections in the denominator.

For all discussed processes, the corrections due to threshold resummation are always positive in the shown mass ranges. The largest $K$-factors occur for $\glu\glu$ production, where in the equal-mass case at $\sqrt{S} = 13.6$~TeV, factors of up to 1.7 can be reached for $\msq = \mgl = 3$~TeV. The corresponding $K$-factors for the decoupled scenario of $\glu\glu$ are a bit smaller and reach only a factor of about 1.5 for $\sqrt{S} = 13.6$~TeV. The lowest $K$-factors for all set-ups occur for $\squ\squ$ production where the effects due to threshold resummation are similar for all energies and increase the NLO cross section only by approximately 10--20\%. The behaviour of the $K$-factor for $\squ\squ^*$ production at $\sqrt{S} = 13$ TeV for masses above 2.5 GeV in the upper left panel of Fig.~\ref{fig:kfactor} has been discussed in \cite{Beenakker:2016lwe}. It is due to setting negative cross sections as obtained for several replicas of the PDF4LHC15 set to zero to obtain a positive central cross section prediction, and the effect is most pronounced for the case of $\squ\squ^*$ production.

%% file: 6-conclusions.tex
%%%%%%%%%%%%%%%%%%%%%%%%%%%%%%%%%%%%%%%%%%%
\section{Conclusions and outlook}\label{s:conclusion}
%%%%%%%%%%%%%%%%%%%%%%%%%%%%%%%%%%%%%%%%%%%
In this work, we report on the update of the predictions for coloured sparticle production at the LHC Run 3 for a collision energy of $\sqrt{S} = 13.6$~TeV using the updated PDF4LHC21 set. The predictions for the total cross sections of the processes of gluino-pair, squark-antisquark, squark-gluino, squark-pair as well as stop-antistop production were computed at \nnlonnll{} accuracy including corrections from the threshold resummation of soft and Coulomb gluons as well as bound-state corrections. To date, the results constitute the state-of-the art theoretical predictions for these types of processes and are used by the experimental ATLAS and CMS collaborations for their analyses of squark and gluino searches. We furthermore describe the update to version 2.0 of the code package \nnllfast{} which includes the new predictions as numerical grids. 
Theoretical uncertainties are, as usual for higher-precision results, found to be reduced significantly compared to the fixed-order calculation at NLO-QCD, with the uncertainty now being dominated by the PDF error for most of the mass ranges.

The \nnllfast{}~2.0 predictions supersede those obtained with \nnllfast{}~1.1, which were computed at 13 TeV using the PDF4LHC15 set. By comparing the predictions, we found that the total cross sections increased uniformly by up to about 60--70\% in the probed mass regions for all processes by changing the centre-of-mass energy from 13~TeV to 13.6~TeV. The update from PDF4LHC15 to the newer PDF4LHC21 set influences the region of heavy squark and gluino masses such that the total effect on the ratio between the \nnllfast{}~2.0 and 1.1 results can reach a factor of above 2, and for some processes even a factor of above 4. We tested that for a future centre-of-mass energy of $\sqrt S = 14$~TeV, the increase as compared to \nnllfast{}~1.1 for $\sqrt S = 13$~TeV could reach even higher factors of up to 6, which could turn out to be relevant for future precise determinations of the mass exclusion limits for squarks and gluinos in case of null results from SUSY searches, or, should a signal for beyond the Standard Model physics compatible with SUSY be found, to study the properties of the new particles.

The updated predictions are made available in form of numerical grids together with an interpolation code on the website of the \nnllfast{} project:
\begin{center}
	\url{https://www.uni-muenster.de/Physik.TP/~akule_01/nnllfast}
\end{center}
Furthermore, for several simplified scenarios, the total cross section numbers for coloured sparticle production together with their uncertainties are available for $\sqrt S = 13.6$~TeV and previous collision energies, amongst other sparticle production processes, on the TWiki page of the LHC SUSY Cross Section Working Group%
\footnote{The predictions for coloured sparticles provided on the TWiki page are calculated as average cross sections $\sigma_{\text{avg}} := (U + L)/2$ with symmetric uncertainties $\delta_{\text{sym}} := (U - L)/(2\sigma_{\text{avg}}) = (U - L)/(U + L)$ with $U$ and $L$ as defined in Eq.~\eqref{eq:upperlowerlimits}, in agreement with the approach of \cite{Borschensky:2014cia}.}%
:
\begin{center}
	\url{https://twiki.cern.ch/twiki/bin/view/LHCPhysics/SUSYCrossSections}
\end{center}

%% file: nnllfast-v2.bbl
\begin{thebibliography}{10}
\providecommand{\url}[1]{{#1}}
\providecommand{\urlprefix}{URL }
\expandafter\ifx\csname urlstyle\endcsname\relax
  \providecommand{\doi}[1]{DOI \discretionary{}{}{}#1}\else
  \providecommand{\doi}{DOI \discretionary{}{}{}\begingroup
  \urlstyle{rm}\Url}\fi

\bibitem{Wess:1973kz}
J.~Wess, B.~Zumino, \emph{{A Lagrangian Model Invariant Under Supergauge
  Transformations}}.
\newblock \href{https://doi.org/10.1016/0370-2693(74)90578-4}{Phys. Lett. B
  \textbf{49}, 52 (1974)}

\bibitem{Wess:1974tw}
J.~Wess, B.~Zumino, \emph{{Supergauge Transformations in Four-Dimensions}}.
\newblock \href{https://doi.org/10.1016/0550-3213(74)90355-1}{Nucl. Phys. B
  \textbf{70}, 39 (1974)}

\bibitem{Fayet:1976et}
P.~Fayet, \emph{{Supersymmetry and Weak, Electromagnetic and Strong
  Interactions}}.
\newblock \href{https://doi.org/10.1016/0370-2693(76)90319-1}{Phys. Lett. B
  \textbf{64}, 159 (1976)}

\bibitem{Farrar:1978xj}
G.R. Farrar, P.~Fayet, \emph{{Phenomenology of the Production, Decay, and
  Detection of New Hadronic States Associated with Supersymmetry}}.
\newblock \href{https://doi.org/10.1016/0370-2693(78)90858-4}{Phys. Lett. B
  \textbf{76}, 575 (1978)}

\bibitem{Sohnius:1985qm}
M.F. Sohnius, \emph{{Introducing Supersymmetry}}.
\newblock \href{https://doi.org/10.1016/0370-1573(85)90023-7}{Phys. Rept.
  \textbf{128}, 39 (1985)}

\bibitem{Martin:1997ns}
S.P. Martin, \emph{{A Supersymmetry primer}}.
\newblock \href{https://doi.org/10.1142/9789812839657_0001}{Adv. Ser. Direct.
  High Energy Phys. \textbf{18}, 1 (1998)},
  \href{https://arxiv.org/abs/hep-ph/9709356}{arXiv:hep-ph/9709356}

\bibitem{Nilles:1983ge}
H.P. Nilles, \emph{{Supersymmetry, Supergravity and Particle Physics}}.
\newblock \href{https://doi.org/10.1016/0370-1573(84)90008-5}{Phys. Rept.
  \textbf{110}, 1 (1984)}

\bibitem{Haber:1984rc}
H.E. Haber, G.L. Kane, \emph{{The Search for Supersymmetry: Probing Physics
  Beyond the Standard Model}}.
\newblock \href{https://doi.org/10.1016/0370-1573(85)90051-1}{Phys. Rept.
  \textbf{117}, 75 (1985)}

\bibitem{ATLAS:1999uwa}
A.~Airapetian, et~al., ATLAS collaboration, \emph{{ATLAS: Detector and physics
  performance technical design report. Volume 1}}.
\newblock CERN-LHCC-99-14, ATLAS-TDR-14

\bibitem{CMS:2006myw}
G.L. Bayatian, et~al., CMS collaboration, \emph{{CMS Physics}: {Technical
  Design Report Volume 1: Detector Performance and Software}}.
\newblock CERN-LHCC-2006-001, CMS-TDR-8-1, CERN-LHCC-2006-001, CMS-TDR-8-1

\bibitem{ATLAS:2017tmw}
M.~Aaboud, et~al., ATLAS collaboration, \emph{{Search for supersymmetry in
  final states with two same-sign or three leptons and jets using 36 fb$^{-1}$
  of $\sqrt{s}=13$ TeV $pp$ collision data with the ATLAS detector}}.
\newblock \href{https://doi.org/10.1007/JHEP09(2017)084}{JHEP \textbf{09}, 084
  (2017)}, \href{https://arxiv.org/abs/1706.03731}{arXiv:1706.03731} [Erratum:
  JHEP 08, 121 (2019)]

\bibitem{ATLAS:2018nud}
M.~Aaboud, et~al., ATLAS collaboration, \emph{{Search for photonic signatures
  of gauge-mediated supersymmetry in 13 TeV $pp$ collisions with the ATLAS
  detector}}.
\newblock \href{https://doi.org/10.1103/PhysRevD.97.092006}{Phys. Rev. D
  \textbf{97}(9), 092006 (2018)},
  \href{https://arxiv.org/abs/1802.03158}{arXiv:1802.03158}

\bibitem{ATLAS:2020syg}
G.~Aad, et~al., ATLAS collaboration, \emph{{Search for squarks and gluinos in
  final states with jets and missing transverse momentum using 139 fb$^{-1}$ of
  $\sqrt{s}$ =13 TeV $pp$ collision data with the ATLAS detector}}.
\newblock \href{https://doi.org/10.1007/JHEP02(2021)143}{JHEP \textbf{02}, 143
  (2021)}, \href{https://arxiv.org/abs/2010.14293}{arXiv:2010.14293}

\bibitem{ATLAS:2021kxv}
G.~Aad, et~al., ATLAS collaboration, \emph{{Search for new phenomena in events
  with an energetic jet and missing transverse momentum in $pp$ collisions at
  $\sqrt {s}$ =13 TeV with the ATLAS detector}}.
\newblock \href{https://doi.org/10.1103/PhysRevD.103.112006}{Phys. Rev. D
  \textbf{103}(11), 112006 (2021)},
  \href{https://arxiv.org/abs/2102.10874}{arXiv:2102.10874}

\bibitem{ATLAS:2022ckd}
G.~Aad, et~al., ATLAS collaboration, \emph{{Search for new phenomena in final
  states with photons, jets and missing transverse momentum in pp collisions at
  $ \sqrt{s} $ = 13 TeV with the ATLAS detector}}.
\newblock \href{https://doi.org/10.1007/JHEP07(2023)021}{JHEP \textbf{07}, 021
  (2023)}, \href{https://arxiv.org/abs/2206.06012}{arXiv:2206.06012}

\bibitem{ATLAS:2022ihe}
G.~Aad, et~al., ATLAS collaboration, \emph{{Search for supersymmetry in final
  states with missing transverse momentum and three or more b-jets in 139
  fb$^{-1}$ of proton\textendash{}proton collisions at $\sqrt{s} = 13$~TeV with
  the ATLAS detector}}.
\newblock \href{https://doi.org/10.1140/epjc/s10052-023-11543-6}{Eur. Phys. J.
  C \textbf{83}(7), 561 (2023)},
  \href{https://arxiv.org/abs/2211.08028}{arXiv:2211.08028}

\bibitem{ATLAS:2023afl}
G.~Aad, et~al., ATLAS collaboration, \emph{{Search for pair production of
  squarks or gluinos decaying via sleptons or weak bosons in final states with
  two same-sign or three leptons with the ATLAS detector}}.
\newblock \href{https://arxiv.org/abs/2307.01094}{arXiv:2307.01094}

\bibitem{CMS:2019zmd}
CMS collaboration, \emph{{Search for supersymmetry in proton-proton collisions
  at 13 TeV in final states with jets and missing transverse momentum}}.
\newblock \href{https://doi.org/10.1007/JHEP10(2019)244}{JHEP \textbf{10}, 244
  (2019)}, \href{https://arxiv.org/abs/1908.04722}{arXiv:1908.04722}

\bibitem{CMS:2019ybf}
A.M. Sirunyan, et~al., CMS collaboration, \emph{{Searches for physics beyond
  the standard model with the $M_\mathrm{T2}$ variable in hadronic final states
  with and without disappearing tracks in proton-proton collisions at
  $\sqrt{s}=$ 13 TeV}}.
\newblock \href{https://doi.org/10.1140/epjc/s10052-019-7493-x}{Eur. Phys. J. C
  \textbf{80}(1), 3 (2020)},
  \href{https://arxiv.org/abs/1909.03460}{arXiv:1909.03460}

\bibitem{CMS:2020cur}
A.M. Sirunyan, et~al., CMS collaboration, \emph{{Search for supersymmetry in pp
  collisions at $\sqrt{s}=$ 13 TeV with 137 fb$^{-1}$ in final states with a
  single lepton using the sum of masses of large-radius jets}}.
\newblock \href{https://doi.org/10.1103/PhysRevD.101.052010}{Phys. Rev. D
  \textbf{101}(5), 052010 (2020)},
  \href{https://arxiv.org/abs/1911.07558}{arXiv:1911.07558}

\bibitem{CMS:2020bfa}
A.M. Sirunyan, et~al., CMS collaboration, \emph{{Search for supersymmetry in
  final states with two oppositely charged same-flavor leptons and missing
  transverse momentum in proton-proton collisions at $\sqrt{s} =$ 13 TeV}}.
\newblock \href{https://doi.org/10.1007/JHEP04(2021)123}{JHEP \textbf{04}, 123
  (2021)}, \href{https://arxiv.org/abs/2012.08600}{arXiv:2012.08600}

\bibitem{CMS:2021beq}
A.M. Sirunyan, et~al., CMS collaboration, \emph{{Search for top squark
  production in fully-hadronic final states in proton-proton collisions at
  $\sqrt{s} =$ 13 TeV}}.
\newblock \href{https://doi.org/10.1103/PhysRevD.104.052001}{Phys. Rev. D
  \textbf{104}(5), 052001 (2021)},
  \href{https://arxiv.org/abs/2103.01290}{arXiv:2103.01290}

\bibitem{CMS:2023xlp}
A.~Hayrapetyan, et~al., CMS collaboration, \emph{{Search for new physics in
  multijet events with at least one photon and large missing transverse
  momentum in proton-proton collisions at 13 TeV}}.
\newblock \href{https://doi.org/10.1007/JHEP10(2023)046}{JHEP \textbf{10}, 046
  (2023)}, \href{https://arxiv.org/abs/2307.16216}{arXiv:2307.16216}

\bibitem{CMS:2023zuu}
A.~Hayrapetyan, et~al., CMS collaboration, \emph{{Search for stealth
  supersymmetry in final states with two photons, jets, and low missing
  transverse momentum in proton-proton collisions at $\sqrt{s}$ = 13 TeV}}.
\newblock \href{https://arxiv.org/abs/2310.03154}{arXiv:2310.03154}

\bibitem{ATLAS:2019gdh}
G.~Aad, et~al., ATLAS collaboration, \emph{{Search for bottom-squark pair
  production with the ATLAS detector in final states containing Higgs bosons,
  $b$-jets and missing transverse momentum}}.
\newblock \href{https://doi.org/10.1007/JHEP12(2019)060}{JHEP \textbf{12}, 060
  (2019)}, \href{https://arxiv.org/abs/1908.03122}{arXiv:1908.03122}

\bibitem{ATLAS:2020dsf}
G.~Aad, et~al., ATLAS collaboration, \emph{{Search for a scalar partner of the
  top quark in the all-hadronic $t{\bar{t}}$ plus missing transverse momentum
  final state at $\sqrt{s}=13$ TeV with the ATLAS detector}}.
\newblock \href{https://doi.org/10.1140/epjc/s10052-020-8102-8}{Eur. Phys. J. C
  \textbf{80}(8), 737 (2020)},
  \href{https://arxiv.org/abs/2004.14060}{arXiv:2004.14060}

\bibitem{ATLAS:2020xzu}
G.~Aad, et~al., ATLAS collaboration, \emph{{Search for new phenomena with top
  quark pairs in final states with one lepton, jets, and missing transverse
  momentum in $pp$ collisions at $ \sqrt{s} $ = 13 TeV with the ATLAS
  detector}}.
\newblock \href{https://doi.org/10.1007/JHEP04(2021)174}{JHEP \textbf{04}, 174
  (2021)}, \href{https://arxiv.org/abs/2012.03799}{arXiv:2012.03799}

\bibitem{ATLAS:2021yij}
G.~Aad, et~al., ATLAS collaboration, \emph{{Search for new phenomena in final
  states with $b$-jets and missing transverse momentum in $\sqrt{s}=13$ TeV
  $pp$ collisions with the ATLAS detector}}.
\newblock \href{https://doi.org/10.1007/JHEP05(2021)093}{JHEP \textbf{05}, 093
  (2021)}, \href{https://arxiv.org/abs/2101.12527}{arXiv:2101.12527}

\bibitem{ATLAS:2021pzz}
G.~Aad, et~al., ATLAS collaboration, \emph{{Search for bottom-squark pair
  production in $pp$ collision events at $\sqrt{s} = 13$ TeV with hadronically
  decaying $\tau$-leptons, $b$-jets and missing transverse momentum using the
  ATLAS detector}}.
\newblock \href{https://doi.org/10.1103/PhysRevD.104.032014}{Phys. Rev. D
  \textbf{104}(3), 032014 (2021)},
  \href{https://arxiv.org/abs/2103.08189}{arXiv:2103.08189}

\bibitem{ATLAS:2021jyv}
G.~Aad, et~al., ATLAS collaboration, \emph{{Search for new phenomena in $pp$
  collisions in final states with tau leptons, b-jets, and missing transverse
  momentum with the ATLAS detector}}.
\newblock \href{https://doi.org/10.1103/PhysRevD.104.112005}{Phys. Rev. D
  \textbf{104}(11), 112005 (2021)},
  \href{https://arxiv.org/abs/2108.07665}{arXiv:2108.07665}

\bibitem{ATLAS:2023dbq}
ATLAS collaboration, \emph{{Search for new phenomena with top-quark pairs in
  final states with one lepton, jets and missing transverse momentum using 140
  $\mathrm{fb}^{-1}$ of data at $\sqrt{s}=13$ TeV with the ATLAS detector}}.
\newblock ATLAS-CONF-2023-043

\bibitem{CMS:2021eha}
A.~Tumasyan, et~al., CMS collaboration, \emph{{Combined searches for the
  production of supersymmetric top quark partners in proton\textendash{}proton
  collisions at $\sqrt{s} = 13\,\text {Te}\text {V} $}}.
\newblock \href{https://doi.org/10.1140/epjc/s10052-021-09721-5}{Eur. Phys. J.
  C \textbf{81}(11), 970 (2021)},
  \href{https://arxiv.org/abs/2107.10892}{arXiv:2107.10892}

\bibitem{CMS:2023ktc}
A.~Tumasyan, et~al., CMS collaboration, \emph{{Search for top squarks in the
  four-body decay mode with single lepton final states in proton-proton
  collisions at $ \sqrt{s} $ = 13 TeV}}.
\newblock \href{https://doi.org/10.1007/JHEP06(2023)060}{JHEP \textbf{06}, 060
  (2023)}, \href{https://arxiv.org/abs/2301.08096}{arXiv:2301.08096}

\bibitem{CMS:2023yzg}
A.~Tumasyan, et~al., CMS collaboration, \emph{{Search for top squark pair
  production in a final state with at least one hadronically decaying tau
  lepton in proton-proton collisions at $ \sqrt{s} $ = 13 TeV}}.
\newblock \href{https://doi.org/10.1007/JHEP07(2023)110}{JHEP \textbf{07}, 110
  (2023)}, \href{https://arxiv.org/abs/2304.07174}{arXiv:2304.07174}

\bibitem{Beenakker:1994an}
W.~Beenakker, R.~Hopker, M.~Spira, P.M. Zerwas, \emph{{Squark production at the
  Tevatron}}.
\newblock \href{https://doi.org/10.1103/PhysRevLett.74.2905}{Phys. Rev. Lett.
  \textbf{74}, 2905 (1995)},
  \href{https://arxiv.org/abs/hep-ph/9412272}{arXiv:hep-ph/9412272}

\bibitem{Beenakker:1995fp}
W.~Beenakker, R.~Hopker, M.~Spira, P.M. Zerwas, \emph{{Gluino pair production
  at the Tevatron}}.
\newblock \href{https://doi.org/10.1007/s002880050016}{Z. Phys. C \textbf{69},
  163 (1995)},
  \href{https://arxiv.org/abs/hep-ph/9505416}{arXiv:hep-ph/9505416}

\bibitem{Beenakker:1996ch}
W.~Beenakker, R.~Hopker, M.~Spira, P.M. Zerwas, \emph{{Squark and gluino
  production at hadron colliders}}.
\newblock \href{https://doi.org/10.1016/S0550-3213(97)80027-2}{Nucl. Phys. B
  \textbf{492}, 51 (1997)},
  \href{https://arxiv.org/abs/hep-ph/9610490}{arXiv:hep-ph/9610490}

\bibitem{Beenakker:1997ut}
W.~Beenakker, M.~Kramer, T.~Plehn, M.~Spira, P.M. Zerwas, \emph{{Stop
  production at hadron colliders}}.
\newblock \href{https://doi.org/10.1016/S0550-3213(98)00014-5}{Nucl. Phys. B
  \textbf{515}, 3 (1998)},
  \href{https://arxiv.org/abs/hep-ph/9710451}{arXiv:hep-ph/9710451}

\bibitem{Hollik:2012rc}
W.~Hollik, J.M. Lindert, D.~Pagani, \emph{{NLO corrections to squark-squark
  production and decay at the LHC}}.
\newblock \href{https://doi.org/10.1007/JHEP03(2013)139}{JHEP \textbf{03}, 139
  (2013)}, \href{https://arxiv.org/abs/1207.1071}{arXiv:1207.1071}

\bibitem{Goncalves-Netto:2012nvl}
D.~Gon\c{c}alves-Netto, D.~L\'opez-Val, K.~Mawatari, T.~Plehn, I.~Wigmore,
  \emph{{Automated Squark and Gluino Production to Next-to-Leading Order}}.
\newblock \href{https://doi.org/10.1103/PhysRevD.87.014002}{Phys. Rev. D
  \textbf{87}(1), 014002 (2013)},
  \href{https://arxiv.org/abs/1211.0286}{arXiv:1211.0286}

\bibitem{Hollik:2013xwa}
W.~Hollik, J.M. Lindert, D.~Pagani, \emph{{On cascade decays of squarks at the
  LHC in NLO QCD}}.
\newblock \href{https://doi.org/10.1140/epjc/s10052-013-2410-1}{Eur. Phys. J. C
  \textbf{73}, 2410 (2013)},
  \href{https://arxiv.org/abs/1303.0186}{arXiv:1303.0186}

\bibitem{Gavin:2013kga}
R.~Gavin, C.~Hangst, M.~Kr\"amer, M.~M\"uhlleitner, M.~Pellen, E.~Popenda,
  M.~Spira, \emph{{Matching Squark Pair Production at NLO with Parton
  Showers}}.
\newblock \href{https://doi.org/10.1007/JHEP10(2013)187}{JHEP \textbf{10}, 187
  (2013)}, \href{https://arxiv.org/abs/1305.4061}{arXiv:1305.4061}

\bibitem{Gavin:2014yga}
R.~Gavin, C.~Hangst, M.~Kr\"amer, M.~M\"uhlleitner, M.~Pellen, E.~Popenda,
  M.~Spira, \emph{{Squark Production and Decay matched with Parton Showers at
  NLO}}.
\newblock \href{https://doi.org/10.1140/epjc/s10052-014-3243-2}{Eur. Phys. J. C
  \textbf{75}(1), 29 (2015)},
  \href{https://arxiv.org/abs/1407.7971}{arXiv:1407.7971}

\bibitem{Degrande:2015vaa}
C.~Degrande, B.~Fuks, V.~Hirschi, J.~Proudom, H.S. Shao, \emph{{Matching
  next-to-leading order predictions to parton showers in supersymmetric QCD}}.
\newblock \href{https://doi.org/10.1016/j.physletb.2016.01.067}{Phys. Lett. B
  \textbf{755}, 82 (2016)},
  \href{https://arxiv.org/abs/1510.00391}{arXiv:1510.00391}

\bibitem{Frixione:2019fxg}
S.~Frixione, B.~Fuks, V.~Hirschi, K.~Mawatari, H.S. Shao, P.A. Sunder, M.~Zaro,
  \emph{{Automated simulations beyond the Standard Model: supersymmetry}}.
\newblock \href{https://doi.org/10.1007/JHEP12(2019)008}{JHEP \textbf{12}, 008
  (2019)}, \href{https://arxiv.org/abs/1907.04898}{arXiv:1907.04898}

\bibitem{Hollik:2007wf}
W.~Hollik, M.~Kollar, M.K. Trenkel, \emph{{Hadronic production of top-squark
  pairs with electroweak NLO contributions}}.
\newblock \href{https://doi.org/10.1088/1126-6708/2008/02/018}{JHEP
  \textbf{02}, 018 (2008)},
  \href{https://arxiv.org/abs/0712.0287}{arXiv:0712.0287}

\bibitem{Beccaria:2008mi}
M.~Beccaria, G.~Macorini, L.~Panizzi, F.M. Renard, C.~Verzegnassi,
  \emph{{Stop-antistop and sbottom-antisbottom production at LHC: A One-loop
  search for model parameters dependence}}.
\newblock \href{https://doi.org/10.1142/S0217751X08041694}{Int. J. Mod. Phys. A
  \textbf{23}, 4779 (2008)},
  \href{https://arxiv.org/abs/0804.1252}{arXiv:0804.1252}

\bibitem{Hollik:2008yi}
W.~Hollik, E.~Mirabella, \emph{{Squark anti-squark pair production at the LHC:
  The Electroweak contribution}}.
\newblock \href{https://doi.org/10.1088/1126-6708/2008/12/087}{JHEP
  \textbf{12}, 087 (2008)},
  \href{https://arxiv.org/abs/0806.1433}{arXiv:0806.1433}

\bibitem{Hollik:2008vm}
W.~Hollik, E.~Mirabella, M.K. Trenkel, \emph{{Electroweak contributions to
  squark-gluino production at the LHC}}.
\newblock \href{https://doi.org/10.1088/1126-6708/2009/02/002}{JHEP
  \textbf{02}, 002 (2009)},
  \href{https://arxiv.org/abs/0810.1044}{arXiv:0810.1044}

\bibitem{Mirabella:2009ap}
E.~Mirabella, \emph{{NLO electroweak contributions to gluino pair production at
  hadron colliders}}.
\newblock \href{https://doi.org/10.1088/1126-6708/2009/12/012}{JHEP
  \textbf{12}, 012 (2009)},
  \href{https://arxiv.org/abs/0908.3318}{arXiv:0908.3318}

\bibitem{Germer:2010vn}
J.~Germer, W.~Hollik, E.~Mirabella, M.K. Trenkel, \emph{{Hadronic production of
  squark-squark pairs: The electroweak contributions}}.
\newblock \href{https://doi.org/10.1007/JHEP08(2010)023}{JHEP \textbf{08}, 023
  (2010)}, \href{https://arxiv.org/abs/1004.2621}{arXiv:1004.2621}

\bibitem{Germer:2014jpa}
J.~Germer, W.~Hollik, J.M. Lindert, E.~Mirabella, \emph{{Top-squark pair
  production at the LHC: a complete analysis at next-to-leading order}}.
\newblock \href{https://doi.org/10.1007/JHEP09(2014)022}{JHEP \textbf{09}, 022
  (2014)}, \href{https://arxiv.org/abs/1404.5572}{arXiv:1404.5572}

\bibitem{Hollik:2015lha}
W.~Hollik, J.M. Lindert, E.~Mirabella, D.~Pagani, \emph{{Electroweak
  corrections to squark-antisquark production at the LHC}}.
\newblock \href{https://doi.org/10.1007/JHEP08(2015)099}{JHEP \textbf{08}, 099
  (2015)}, \href{https://arxiv.org/abs/1506.01052}{arXiv:1506.01052}

\bibitem{Sterman:1986aj}
G.F. Sterman, \emph{{Summation of Large Corrections to Short Distance Hadronic
  Cross-Sections}}.
\newblock \href{https://doi.org/10.1016/0550-3213(87)90258-6}{Nucl. Phys. B
  \textbf{281}, 310 (1987)}

\bibitem{Catani:1989ne}
S.~Catani, L.~Trentadue, \emph{{Resummation of the QCD Perturbative Series for
  Hard Processes}}.
\newblock \href{https://doi.org/10.1016/0550-3213(89)90273-3}{Nucl. Phys. B
  \textbf{327}, 323 (1989)}

\bibitem{Bonciani:1998vc}
R.~Bonciani, S.~Catani, M.L. Mangano, P.~Nason, \emph{{NLL resummation of the
  heavy quark hadroproduction cross-section}}.
\newblock \href{https://doi.org/10.1016/S0550-3213(98)00335-6}{Nucl. Phys. B
  \textbf{529}, 424 (1998)},
  \href{https://arxiv.org/abs/hep-ph/9801375}{arXiv:hep-ph/9801375} [Erratum:
  Nucl.Phys.B 803, 234 (2008)]

\bibitem{Contopanagos:1996nh}
H.~Contopanagos, E.~Laenen, G.F. Sterman, \emph{{Sudakov factorization and
  resummation}}.
\newblock \href{https://doi.org/10.1016/S0550-3213(96)00567-6}{Nucl. Phys. B
  \textbf{484}, 303 (1997)},
  \href{https://arxiv.org/abs/hep-ph/9604313}{arXiv:hep-ph/9604313}

\bibitem{Kidonakis:1998bk}
N.~Kidonakis, G.~Oderda, G.F. Sterman, \emph{{Threshold resummation for dijet
  cross-sections}}.
\newblock \href{https://doi.org/10.1016/S0550-3213(98)00243-0}{Nucl. Phys. B
  \textbf{525}, 299 (1998)},
  \href{https://arxiv.org/abs/hep-ph/9801268}{arXiv:hep-ph/9801268}

\bibitem{Kidonakis:1998nf}
N.~Kidonakis, G.~Oderda, G.F. Sterman, \emph{{Evolution of color exchange in
  QCD hard scattering}}.
\newblock \href{https://doi.org/10.1016/S0550-3213(98)00441-6}{Nucl. Phys. B
  \textbf{531}, 365 (1998)},
  \href{https://arxiv.org/abs/hep-ph/9803241}{arXiv:hep-ph/9803241}

\bibitem{Kulesza:2008jb}
A.~Kulesza, L.~Motyka, \emph{{Threshold resummation for squark-antisquark and
  gluino-pair production at the LHC}}.
\newblock \href{https://doi.org/10.1103/PhysRevLett.102.111802}{Phys. Rev.
  Lett. \textbf{102}, 111802 (2009)},
  \href{https://arxiv.org/abs/0807.2405}{arXiv:0807.2405}

\bibitem{Kulesza:2009kq}
A.~Kulesza, L.~Motyka, \emph{{Soft gluon resummation for the production of
  gluino-gluino and squark-antisquark pairs at the LHC}}.
\newblock \href{https://doi.org/10.1103/PhysRevD.80.095004}{Phys. Rev. D
  \textbf{80}, 095004 (2009)},
  \href{https://arxiv.org/abs/0905.4749}{arXiv:0905.4749}

\bibitem{Beenakker:2009ha}
W.~Beenakker, S.~Brensing, M.~Kramer, A.~Kulesza, E.~Laenen, I.~Niessen,
  \emph{{Soft-gluon resummation for squark and gluino hadroproduction}}.
\newblock \href{https://doi.org/10.1088/1126-6708/2009/12/041}{JHEP
  \textbf{12}, 041 (2009)},
  \href{https://arxiv.org/abs/0909.4418}{arXiv:0909.4418}

\bibitem{Beenakker:2010nq}
W.~Beenakker, S.~Brensing, M.~Kramer, A.~Kulesza, E.~Laenen, I.~Niessen,
  \emph{{Supersymmetric top and bottom squark production at hadron colliders}}.
\newblock \href{https://doi.org/10.1007/JHEP08(2010)098}{JHEP \textbf{08}, 098
  (2010)}, \href{https://arxiv.org/abs/1006.4771}{arXiv:1006.4771}

\bibitem{Beenakker:2011fu}
W.~Beenakker, S.~Brensing, M.n. Kramer, A.~Kulesza, E.~Laenen, L.~Motyka,
  I.~Niessen, \emph{{Squark and Gluino Hadroproduction}}.
\newblock \href{https://doi.org/10.1142/S0217751X11053560}{Int. J. Mod. Phys. A
  \textbf{26}, 2637 (2011)},
  \href{https://arxiv.org/abs/1105.1110}{arXiv:1105.1110}

\bibitem{Beenakker:2011dk}
W.~Beenakker, S.~Brensing, M.~D'Onofrio, M.~Kramer, A.~Kulesza, E.~Laenen,
  M.~Martinez, I.~Niessen, \emph{{Improved Squark and Gluino Mass Limits from
  Searches for Supersymmetry at Hadron Colliders}}.
\newblock \href{https://doi.org/10.1103/PhysRevD.85.075014}{Phys. Rev. D
  \textbf{85}, 075014 (2012)},
  \href{https://arxiv.org/abs/1106.5647}{arXiv:1106.5647}

\bibitem{Borschensky:2014cia}
C.~Borschensky, M.~Kr\"amer, A.~Kulesza, M.~Mangano, S.~Padhi, T.~Plehn,
  X.~Portell, \emph{{Squark and gluino production cross sections in pp
  collisions at $\sqrt{s}$ = 13, 14, 33 and 100 TeV}}.
\newblock \href{https://doi.org/10.1140/epjc/s10052-014-3174-y}{Eur. Phys. J. C
  \textbf{74}(12), 3174 (2014)},
  \href{https://arxiv.org/abs/1407.5066}{arXiv:1407.5066}

\bibitem{Beenakker:2015rna}
W.~Beenakker, C.~Borschensky, M.~Kr\"amer, A.~Kulesza, E.~Laenen, S.~Marzani,
  J.~Rojo, \emph{{NLO+NLL squark and gluino production cross-sections with
  threshold-improved parton distributions}}.
\newblock \href{https://doi.org/10.1140/epjc/s10052-016-3892-4}{Eur. Phys. J. C
  \textbf{76}(2), 53 (2016)},
  \href{https://arxiv.org/abs/1510.00375}{arXiv:1510.00375}

\bibitem{Beneke:2009rj}
M.~Beneke, P.~Falgari, C.~Schwinn, \emph{{Soft radiation in heavy-particle pair
  production: All-order colour structure and two-loop anomalous dimension}}.
\newblock \href{https://doi.org/10.1016/j.nuclphysb.2009.11.004}{Nucl. Phys. B
  \textbf{828}, 69 (2010)},
  \href{https://arxiv.org/abs/0907.1443}{arXiv:0907.1443}

\bibitem{Beneke:2010da}
M.~Beneke, P.~Falgari, C.~Schwinn, \emph{{Threshold resummation for pair
  production of coloured heavy (s)particles at hadron colliders}}.
\newblock \href{https://doi.org/10.1016/j.nuclphysb.2010.09.009}{Nucl. Phys. B
  \textbf{842}, 414 (2011)},
  \href{https://arxiv.org/abs/1007.5414}{arXiv:1007.5414}

\bibitem{Falgari:2012hx}
P.~Falgari, C.~Schwinn, C.~Wever, \emph{{NLL soft and Coulomb resummation for
  squark and gluino production at the LHC}}.
\newblock \href{https://doi.org/10.1007/JHEP06(2012)052}{JHEP \textbf{06}, 052
  (2012)}, \href{https://arxiv.org/abs/1202.2260}{arXiv:1202.2260}

\bibitem{Beenakker:2011sf}
W.~Beenakker, S.~Brensing, M.~Kramer, A.~Kulesza, E.~Laenen, I.~Niessen,
  \emph{{NNLL resummation for squark-antisquark pair production at the LHC}}.
\newblock \href{https://doi.org/10.1007/JHEP01(2012)076}{JHEP \textbf{01}, 076
  (2012)}, \href{https://arxiv.org/abs/1110.2446}{arXiv:1110.2446}

\bibitem{Langenfeld:2012ti}
U.~Langenfeld, S.O. Moch, T.~Pfoh, \emph{{QCD threshold corrections for gluino
  pair production at hadron colliders}}.
\newblock \href{https://doi.org/10.1007/JHEP11(2012)070}{JHEP \textbf{11}, 070
  (2012)}, \href{https://arxiv.org/abs/1208.4281}{arXiv:1208.4281}

\bibitem{Pfoh:2013iia}
T.~Pfoh, \emph{{Phenomenology of QCD threshold resummation for gluino pair
  production at NNLL}}.
\newblock \href{https://doi.org/10.1007/JHEP05(2013)044}{JHEP \textbf{05}, 044
  (2013)}, \href{https://arxiv.org/abs/1302.7202}{arXiv:1302.7202} [Erratum:
  JHEP 10, 090 (2013)]

\bibitem{Beenakker:2013mva}
W.~Beenakker, T.~Janssen, S.~Lepoeter, M.~Kr\"amer, A.~Kulesza, E.~Laenen,
  I.~Niessen, S.~Thewes, T.~Van~Daal, \emph{{Towards NNLL resummation: hard
  matching coefficients for squark and gluino hadroproduction}}.
\newblock \href{https://doi.org/10.1007/JHEP10(2013)120}{JHEP \textbf{10}, 120
  (2013)}, \href{https://arxiv.org/abs/1304.6354}{arXiv:1304.6354}

\bibitem{Beenakker:2014sma}
W.~Beenakker, C.~Borschensky, M.~Kr\"amer, A.~Kulesza, E.~Laenen, V.~Theeuwes,
  S.~Thewes, \emph{{NNLL resummation for squark and gluino production at the
  LHC}}.
\newblock \href{https://doi.org/10.1007/JHEP12(2014)023}{JHEP \textbf{12}, 023
  (2014)}, \href{https://arxiv.org/abs/1404.3134}{arXiv:1404.3134}

\bibitem{Beenakker:2016gmf}
W.~Beenakker, C.~Borschensky, R.~Heger, M.~Kr\"amer, A.~Kulesza, E.~Laenen,
  \emph{{NNLL resummation for stop pair-production at the LHC}}.
\newblock \href{https://doi.org/10.1007/JHEP05(2016)153}{JHEP \textbf{05}, 153
  (2016)}, \href{https://arxiv.org/abs/1601.02954}{arXiv:1601.02954}

\bibitem{Beenakker:2016lwe}
W.~Beenakker, C.~Borschensky, M.~Kr\"amer, A.~Kulesza, E.~Laenen,
  \emph{{NNLL-fast: predictions for coloured supersymmetric particle production
  at the LHC with threshold and Coulomb resummation}}.
\newblock \href{https://doi.org/10.1007/JHEP12(2016)133}{JHEP \textbf{12}, 133
  (2016)}, \href{https://arxiv.org/abs/1607.07741}{arXiv:1607.07741}

\bibitem{Beneke:2013opa}
M.~Beneke, P.~Falgari, J.~Piclum, C.~Schwinn, C.~Wever, \emph{{Higher-order
  soft and Coulomb corrections to squark and gluino production at the LHC}}.
\newblock \href{https://doi.org/10.22323/1.197.0051}{PoS \textbf{RADCOR2013},
  051 (2013)}, \href{https://arxiv.org/abs/1312.0837}{arXiv:1312.0837}

\bibitem{Broggio:2013cia}
A.~Broggio, A.~Ferroglia, M.~Neubert, L.~Vernazza, L.L. Yang, \emph{{NNLL
  Momentum-Space Resummation for Stop-Pair Production at the LHC}}.
\newblock \href{https://doi.org/10.1007/JHEP03(2014)066}{JHEP \textbf{03}, 066
  (2014)}, \href{https://arxiv.org/abs/1312.4540}{arXiv:1312.4540}

\bibitem{Beneke:2014wda}
M.~Beneke, P.~Falgari, J.~Piclum, C.~Schwinn, C.~Wever, \emph{{Higher-order
  soft and Coulomb corrections to squark and gluino production at the LHC}}.
\newblock \href{https://doi.org/10.22323/1.211.0060}{PoS \textbf{LL2014}, 060
  (2014)}

\bibitem{Beneke:2016kvz}
M.~Beneke, J.~Piclum, C.~Schwinn, C.~Wever, \emph{{NNLL soft and Coulomb
  resummation for squark and gluino production at the LHC}}.
\newblock \href{https://doi.org/10.1007/JHEP10(2016)054}{JHEP \textbf{10}, 054
  (2016)}, \href{https://arxiv.org/abs/1607.07574}{arXiv:1607.07574}

\bibitem{Borschensky:2024zdg}
C.~Borschensky, F.~Frisenna, W.~Kotlarski, A.~Kulesza, D.~St\"ockinger,
  \emph{{Squark production with R-symmetry beyond NLO at the LHC}}.
\newblock \href{https://arxiv.org/abs/2402.10160}{arXiv:2402.10160}

\bibitem{Kribs:2007ac}
G.D. Kribs, E.~Poppitz, N.~Weiner, \emph{{Flavor in supersymmetry with an
  extended R-symmetry}}.
\newblock \href{https://doi.org/10.1103/PhysRevD.78.055010}{Phys. Rev. D
  \textbf{78}, 055010 (2008)},
  \href{https://arxiv.org/abs/0712.2039}{arXiv:0712.2039}

\bibitem{Diessner:2017ske}
P.~Diessner, W.~Kotlarski, S.~Liebschner, D.~St\"ockinger, \emph{{Squark
  production in R-symmetric SUSY with Dirac gluinos: NLO corrections}}.
\newblock \href{https://doi.org/10.3204/PUBDB-2017-10328}{JHEP \textbf{10}, 142
  (2017)}, \href{https://arxiv.org/abs/1707.04557}{arXiv:1707.04557}

\bibitem{Beenakker:1996ed}
W.~Beenakker, R.~Hopker, M.~Spira, \emph{{PROSPINO: A Program for the
  production of supersymmetric particles in next-to-leading order QCD}}.
\newblock \href{https://arxiv.org/abs/hep-ph/9611232}{arXiv:hep-ph/9611232} and
  online at
  \url{https://www.thphys.uni-heidelberg.de/~plehn/index.php?show=prospino} or
  \url{http://tiger.web.psi.ch/prospino/}.

\bibitem{Beneke:2009ye}
M.~Beneke, M.~Czakon, P.~Falgari, A.~Mitov, C.~Schwinn, \emph{{Threshold
  expansion of the $gg(q \bar q)$ $\to \overline {QQ} + X$ cross section at
  $O(\alpha^4_s)$}}.
\newblock \href{https://doi.org/10.1016/j.physletb.2010.05.038}{Phys. Lett. B
  \textbf{690}, 483 (2010)},
  \href{https://arxiv.org/abs/0911.5166}{arXiv:0911.5166} [Erratum: Phys.Lett.B
  778, 464--464 (2018)]

\bibitem{Catani:1996yz}
S.~Catani, M.L. Mangano, P.~Nason, L.~Trentadue, \emph{{The Resummation of soft
  gluons in hadronic collisions}}.
\newblock \href{https://doi.org/10.1016/0550-3213(96)00399-9}{Nucl. Phys. B
  \textbf{478}, 273 (1996)},
  \href{https://arxiv.org/abs/hep-ph/9604351}{arXiv:hep-ph/9604351}

\bibitem{PDF4LHCWorkingGroup:2022cjn}
R.D. Ball, et~al., PDF4LHC Working Group collaboration, \emph{{The PDF4LHC21
  combination of global PDF fits for the LHC Run III}}.
\newblock \href{https://doi.org/10.1088/1361-6471/ac7216}{J. Phys. G
  \textbf{49}(8), 080501 (2022)},
  \href{https://arxiv.org/abs/2203.05506}{arXiv:2203.05506}

\bibitem{Hou:2019efy}
T.J. Hou, et~al., \emph{{New CTEQ global analysis of quantum chromodynamics
  with high-precision data from the LHC}}.
\newblock \href{https://doi.org/10.1103/PhysRevD.103.014013}{Phys. Rev. D
  \textbf{103}(1), 014013 (2021)},
  \href{https://arxiv.org/abs/1912.10053}{arXiv:1912.10053}

\bibitem{Bailey:2020ooq}
S.~Bailey, T.~Cridge, L.A. Harland-Lang, A.D. Martin, R.S. Thorne,
  \emph{{Parton distributions from LHC, HERA, Tevatron and fixed target data:
  MSHT20 PDFs}}.
\newblock \href{https://doi.org/10.1140/epjc/s10052-021-09057-0}{Eur. Phys. J.
  C \textbf{81}(4), 341 (2021)},
  \href{https://arxiv.org/abs/2012.04684}{arXiv:2012.04684}

\bibitem{NNPDF:2017mvq}
R.D. Ball, et~al., NNPDF collaboration, \emph{{Parton distributions from
  high-precision collider data}}.
\newblock \href{https://doi.org/10.1140/epjc/s10052-017-5199-5}{Eur. Phys. J. C
  \textbf{77}(10), 663 (2017)},
  \href{https://arxiv.org/abs/1706.00428}{arXiv:1706.00428}

\bibitem{AbdusSalam:2011fc}
S.S. AbdusSalam, et~al., \emph{{Benchmark Models, Planes, Lines and Points for
  Future SUSY Searches at the LHC}}.
\newblock \href{https://doi.org/10.1140/epjc/s10052-011-1835-7}{Eur. Phys. J. C
  \textbf{71}, 1835 (2011)},
  \href{https://arxiv.org/abs/1109.3859}{arXiv:1109.3859}

\bibitem{Buckley:2014ana}
A.~Buckley, J.~Ferrando, S.~Lloyd, K.~Nordstr\"om, B.~Page, M.~R\"ufenacht,
  M.~Sch\"onherr, G.~Watt, \emph{{LHAPDF6: parton density access in the LHC
  precision era}}.
\newblock \href{https://doi.org/10.1140/epjc/s10052-015-3318-8}{Eur. Phys. J. C
  \textbf{75}, 132 (2015)},
  \href{https://arxiv.org/abs/1412.7420}{arXiv:1412.7420}

\bibitem{Bonvini:2015ira}
M.~Bonvini, S.~Marzani, J.~Rojo, L.~Rottoli, M.~Ubiali, R.D. Ball, V.~Bertone,
  S.~Carrazza, N.P. Hartland, \emph{{Parton distributions with threshold
  resummation}}.
\newblock \href{https://doi.org/10.1007/JHEP09(2015)191}{JHEP \textbf{09}, 191
  (2015)}, \href{https://arxiv.org/abs/1507.01006}{arXiv:1507.01006}

\bibitem{Butterworth:2015oua}
J.~Butterworth, et~al., \emph{{PDF4LHC recommendations for LHC Run II}}.
\newblock \href{https://doi.org/10.1088/0954-3899/43/2/023001}{J. Phys. G
  \textbf{43}, 023001 (2016)},
  \href{https://arxiv.org/abs/1510.03865}{arXiv:1510.03865}

\end{thebibliography}
